\documentclass[fleqn,12pt]{article}
\usepackage[numbers,sort&compress]{natbib}
\usepackage{graphics}
\usepackage{graphicx}
\usepackage{amsmath}
\usepackage{amssymb}
\usepackage{pifont}
\usepackage{amsfonts}
\usepackage{caption2}
\usepackage{float}
\usepackage{color}
\usepackage[a4paper,height=25cm,hmargin={2cm,2cm}]{geometry}
\makeatletter
\newcommand\figcaption{\def\@captype{figure}\caption}
\newcommand\tabcaption{\def\@captype{table}\caption}
\makeatother

\begin{document}
\title{Performing Hong-Ou-Mandel-type Numerical Experiments with
Repulsive Condensates: The case of Dark and Dark-bright Solitons}
\date{(\today)}

\author{Zhi-Yuan Sun\thanks{sunzhiyuan137@aliyun.com}
\\\textit{Department of Physics,} \\\textit{Technion-Israel Institute of
Technology, Haifa, 32000, Israel}
\vspace{5mm}\\
Panayotis\ G.\ Kevrekidis
\\\textit{Department of Mathematics and Statistics, }\\\textit{University of Massachusetts,
Amherst, MA, USA}
\vspace{5mm}\\
Peter Kr\"uger
\\\textit{Midlands Ultracold Atom Research Centre, School of Physics \& Astronomy,}
\\\textit{The University of Nottingham, Nottingham, UK}\\}
\maketitle

\begin{abstract}
The Hong-Ou-Mandel experiment leads indistinguishable photons
simultaneously reaching a 50:50 beam splitter to emerge on the same
port through two-photon interference. Motivated by this phenomenon,
we consider numerical experiments of the same flavor for classical,
{\it wave} objects in the setting of repulsive condensates. We
examine dark solitons interacting with a repulsive barrier, a case
in which we find no significant asymmetries in the emerging waves
after the collision, presumably due to their topological nature. We
also consider case examples of two-component systems, where the dark
solitons trap a bright structure in the second-component
(dark-bright solitary waves). For these, pronounced asymmetries upon
collision {\it are} possible for the non-topological bright
component. We also show an example of a similar phenomenology for
ring dark-bright structures in two dimensions.
\end{abstract}

\newpage
\section{Introduction}
The well-known Hong-Ou-Mandel (HOM) effect in quantum mechanics
describes particle interference of two indistinguishable photons~\cite{Hong}:
when two identical single-photon wave
packets simultaneously enter a 50:50 beam splitter, one in
each input port, both always exit the splitter at the same output
port, although each photon has (on its own) a 50:50 possibility
to exit either output port. With this effect, we can test (by the
manner of the so-called HOM dip)
the degree of indistinguishability of two incoming photons
experimentally. Moreover, the HOM effect has been applied to
demonstrate the purity of a solid-state single-photon
source \cite{Santori} and has provided a
mechanism for logic gates in linear optical quantum computation
\cite{Knill}. Experimental realizations have also been implemented
for larger particle numbers
such as three photons impinging on a multiport mixer~\cite{Campos2000}, and for one and two-photon pairs~\cite{CosmeEtAl08}. Multi-photon experiments and the
associated generalizations of the Hong-Ou-Mandel effect
have been reviewed in~\cite{Ou2007}.

Recent studies have generalized the HOM effect to the interference
of massive particles \cite{Lim,Longo,Laloe,Gertjerenken0,Laloe2}. In fact,
Bose-Einstein condensates (BECs) at very low temperatures provide a
setup for studying an analog to the HOM effect for massive
(bosonic) particles, such as atoms.
Lewis-Swan and Kheruntsyan realized the HOM effect for
massive particles by using a collision of two BECs and a sequence of
laser-induced Bragg pulses as the splitter \cite{Swan}. On the other
hand, this has been further explored experimentally in a plasmonic
setup using surface plasmon polaritons to interact through a
semitransparent Bragg mirror \cite{Martino}.

Considering the more \textit{classical} aspect of
matter waves, solitary waves or solitons have been extensively
studied in the context of BECs; for a recent review, see,
e.g.,~\cite{Kevrekidis}. Bright solitary waves
for attractive interactions have been created in $^7$Li \cite{Khaykovich,
Strecker} and $^{85}$Rb \cite{Cornish}, and their interactions
(also with barriers) have been explored both at the mean-field
and at the quantum-mechanical level \cite{Billam,Helm,Cuevas,Martin,
Gertjerenken,Marchant,Nguyen}. At the junction of the HOM effect and
matter-wave solitons, we previously have proposed a mean-field analogue
of the HOM effect with bright solitons in BECs \cite{Sun}. In our setup,
the bright solitons play the role of a classical wave analogue to the
quantum photons, while the role of the beam splitter is played by a
repulsive Gaussian barrier. Although these are not quantum mechanical
objects at the level of consideration of~\cite{Sun},
our analysis showed that their wave character is responsible
for an intriguing phenomenology.
In particular, we showed that even very slight deviations
of the bright solitons from perfect symmetry (of the order of a few percent
in the relative speed, or in the relative amplitude) yield an
output whereby the bright solitons emerge essentially in only one of the
two ports. This feature is demonstrated to be generic in a wide regime of
soliton and barrier parameters.

It is then natural to inquire whether similar phenomena may be
present in the context of repulsive BECs. While the work of~\cite{Swan}
considered this possibility between two BECs, here we consider
it at the level of topological wave excitations existing within
the (same) BEC.
In particular, we consider the potential of
HOM phenomenology with dark (single or multi-component) solitons.
Dark matter-wave solitons, which are characterized
by localized dips in the atomic density with certain phase slip
across their center, have received considerable research interest in
atomic systems in recent
years~\cite{Frantzeskakis,Kevrekidis}. In BECs they
can be created by phase imprinting~\cite{Burger,Denschlag,Becker},
destructive interference~\cite{oberthaler,oberthaler2},
density engineering~\cite{Anderson}, and by dragging a potential
barrier through the condensate~\cite{Pavloff,Engels,Hans}, among
others.
Collisions of DSs in an elongated BEC have
been observed experimentally \cite{Stellmer,oberthaler,oberthaler2},
showing their potential non-destructive transmission or
reflection with a shift
in their trajectories. However, it is important to caution
here about the necessity for the quasi-one-dimensional
nature of the associated geometry, as under less restrictive
trapping conditions, different types of collisional effects may
arise~\cite{becker2}. On the other hand, interactions of the DSs
with localized impurities have been considered in the literature
\cite{Kivshar0,Konotop}, with relevant investigations proposed also in
the context of BECs \cite{Frantzeskakis1,Bilas,Herring,Tsitoura}.
Moreover, such issues on soliton-impurity interactions have been
extended to dark-bright (DB) solitons \cite{Achilleos,Alvarez},
ring dark solitons \cite{Xue}, and vortices \cite{Davis,Ma}.
While ring dark solitons have yet to be observed as stable objects
experimentally, despite theoretical proposals for their
stabilization~\cite{manjun2}, DB structures have been a focus of
considerable experimental interest, as is evidenced by a
relevant recent review~\cite{revip}.

In our numerical experiments reported here,
we start from dark solitons in repulsive BECs
(within the mean-field description of the quasi-1D
Gross-Pitaevskii (GP) equation with repulsive interactions), and
arrange systematic simulations for the collisions between dark-soliton
pair and impurity. Unlike the bright solitons, we find that
scattering of the dark-soliton pair (with slight asymmetry) by the
impurity is not able to effectively yield the strongly asymmetric
behavior reminiscent of the HOM effect.
Thus, we further pay attention to the DB solitons in a two-component
BEC, with the localized Gaussian impurity (either repulsive or attractive)
added on the bright-soliton component. In such a case, systematic
simulations show that the dark-soliton component presents an analog of
the HOM effect generically. Finally, we give a
prototypical case example of  analogous behavior
in a 2D setup for the ring DB solitons~\cite{stockhofe}.

\section{Scattering of dark-soliton pair by impurity}
Firstly, we examine the collision phenomenology in the setting
of the normalized quasi-1D GP equation with repulsive interactions:
\begin{equation}
i \frac{\partial \psi(x,t)}{\partial t} = \left[ -\frac{1}{2}
\frac{\partial^2}{\partial x^2} + |\psi(x,t)|^2 +
\frac{q}{\sqrt{2\pi} \sigma} e^{-\frac{x^2}{2\sigma^2}} \right]
\psi(x,t)~,\label{1}
\end{equation}
where $\psi(x,t)$ is the dimensionless wave function with normalized
temporal and spatial coordinates $t$ and $x$, and the Gaussian barrier
has a normalized width $\sigma$ and strength $q$. Derivation of the
dimensionless form of this equation, and discussion of the relevant
physical units can be seen, e.g., in \cite{Kevrekidis,Helm,Frantzeskakis1}.
By a procedure similar to \cite{Frantzeskakis1,Kivshar}, we
firstly calculate the profile of the background field with impurity,
$\psi(x,t) = \psi_b(x)e^{-i\psi_0^2 t}$, where $\psi_0^2$ is the normalized
density of the BEC cloud:
\begin{equation}
\psi_b(x) \approx \psi_0 -\frac{q}{4}
e^{2\sigma^2\psi_0^2} \left[ e^{-2\psi_0 x} \textrm{erfc} \left(
\frac{-x+2\sigma^2 \psi_0}{\sqrt{2}\sigma}\right) + e^{2\psi_0 x}
\textrm{erfc} \left( \frac{x+2\sigma^2
\psi_0}{\sqrt{2}\sigma}\right) \right]~,\label{2}
\end{equation}
with the assumption that the impurity is small, where
$\textrm{erfc}(z) = 1-\frac{2}{\sqrt{\pi}}\int_0^z e^{-\eta^2}
d\eta$ gives the complementary error function. The background
field density $\psi_b^2(x)$ describes an effective Thomas-Fermi-like
condensate wave function modified by the localized impurity.
Dynamics of a single dark soliton on top of such a background
with impurity can be approximately described by an adiabatic
perturbation, which is briefly summarized in the appendix.
\\[\intextsep]
\begin{minipage}{\textwidth}
\renewcommand{\captionlabeldelim}{: }
\renewcommand{\figurename}{Figure }
\renewcommand{\captionfont}{ }
\renewcommand{\captionlabelfont}{ }
\vspace{0cm}\centering
\includegraphics[scale=0.65]{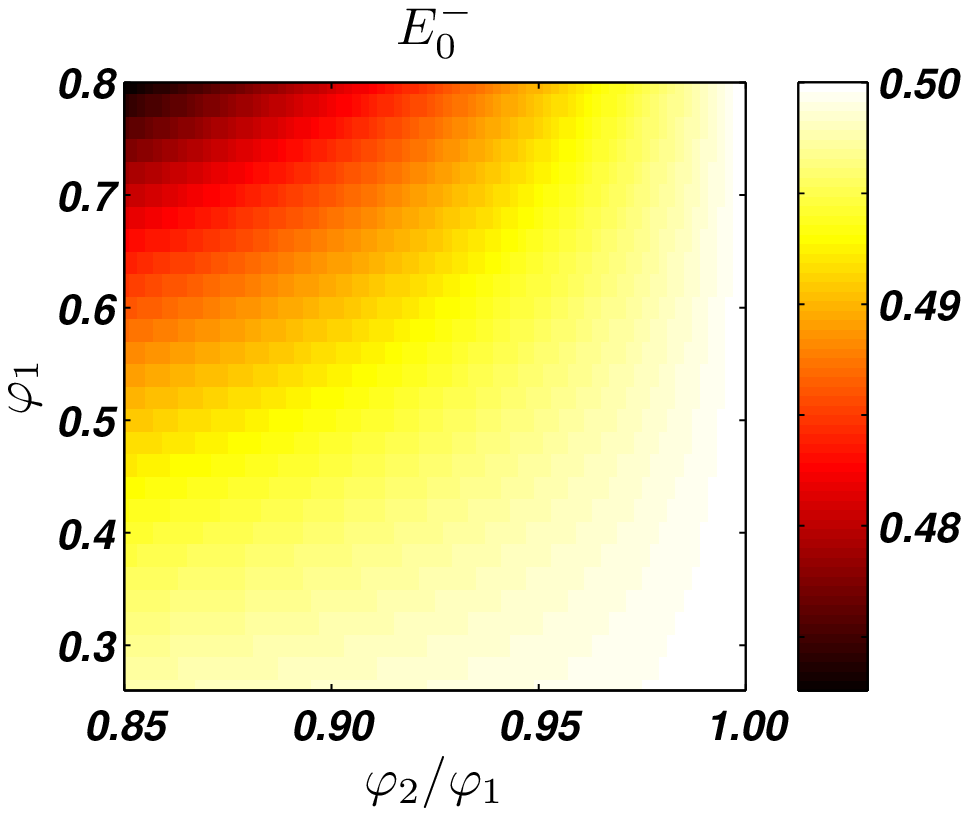}
\includegraphics[scale=0.65]{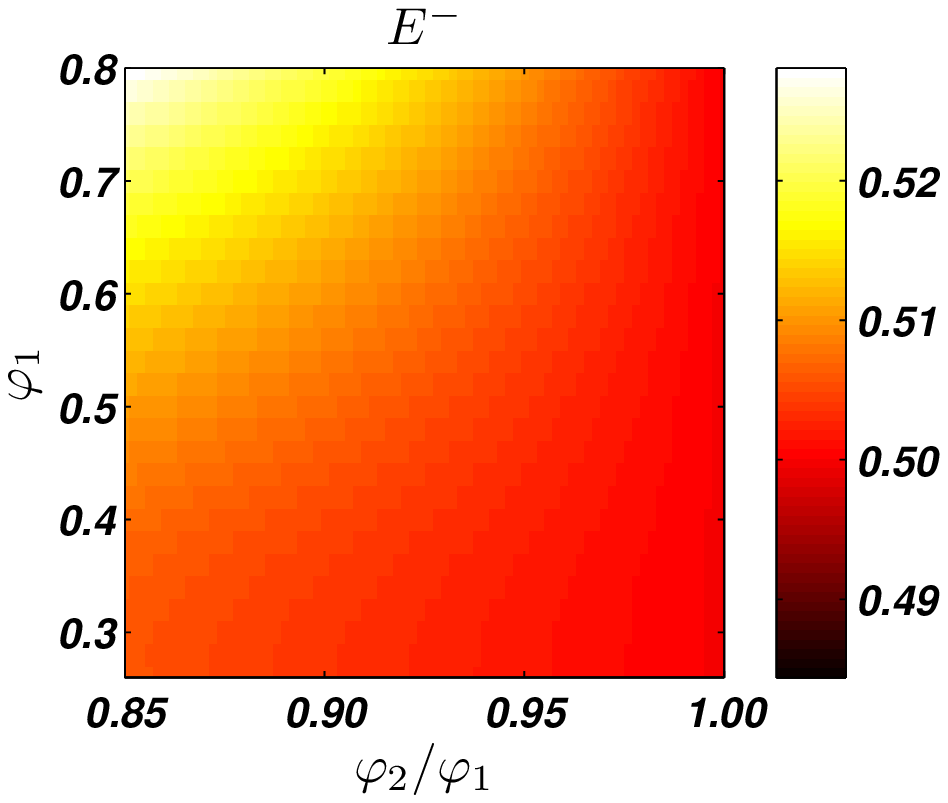}\\
\includegraphics[scale=0.65]{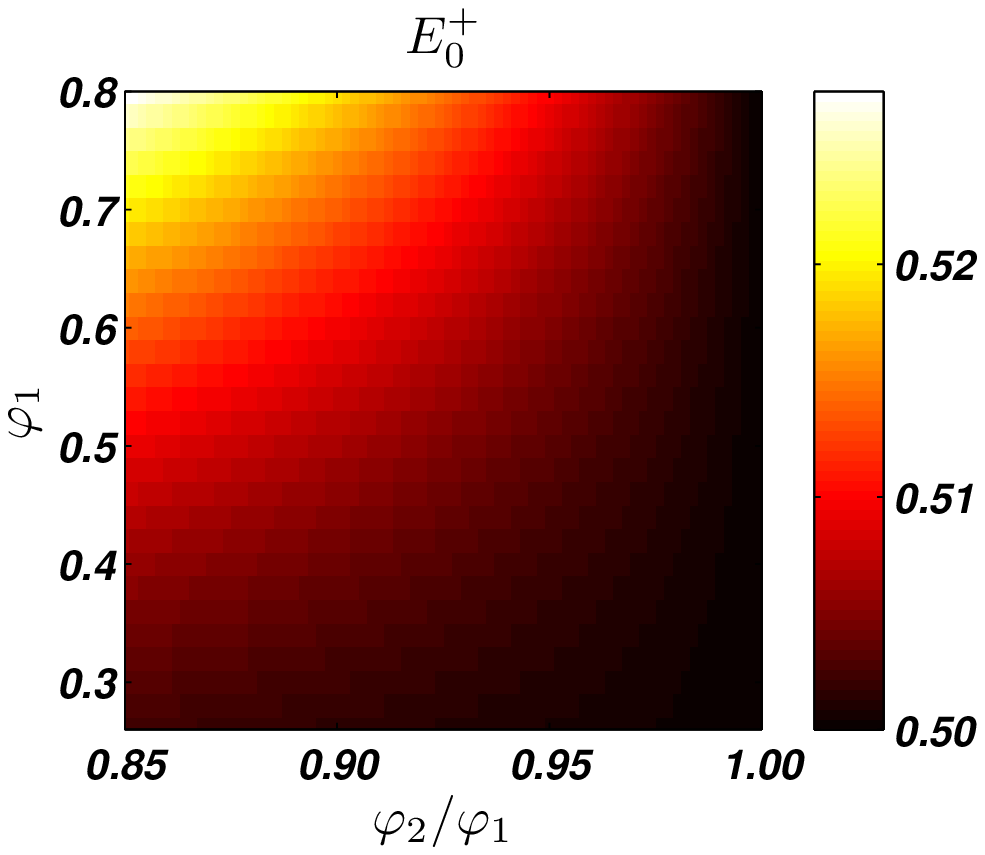}
\includegraphics[scale=0.65]{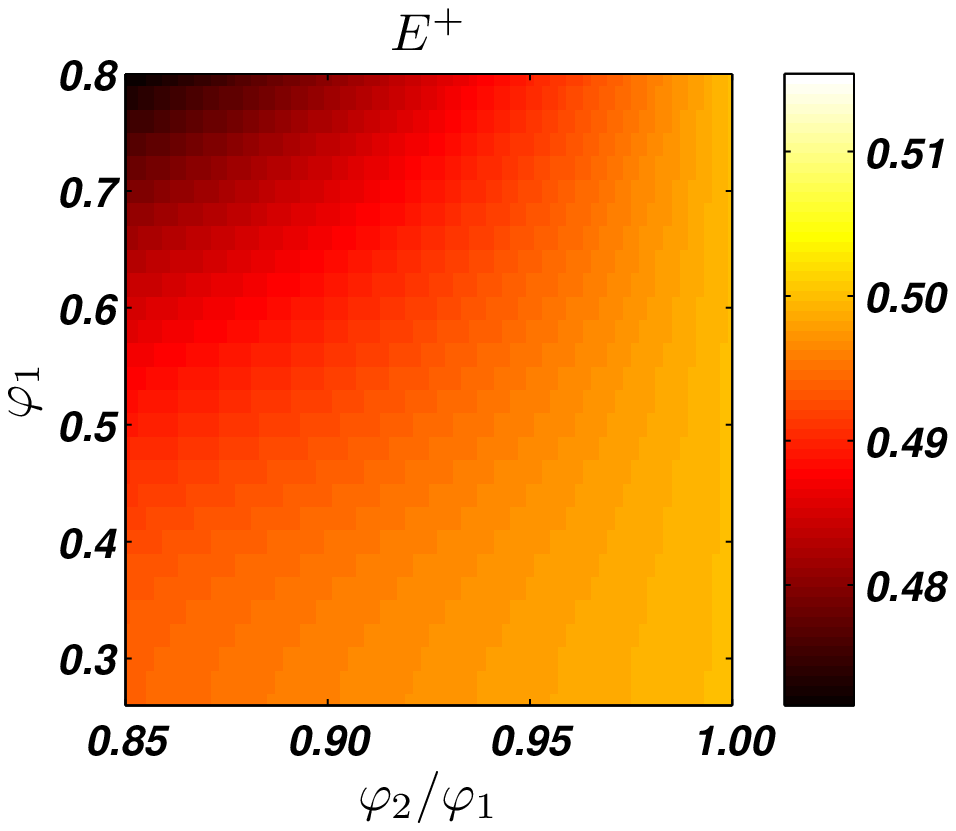}
\figcaption{Phase diagram of $E^\pm$ before (the left two panels for
$t=0$) and after (the right two panels for $t=1.6x_1/\sin\varphi_1$)
collision. The relevant parameters are $q=0.3$, $\sigma=0.1$,
$\psi_0=1$, and $x_1=15$. }\label{A}
\end{minipage}
\\[\intextsep]

For scattering of a dark-soliton pair with small asymmetry,
we perform direct simulations of Eq.~(\ref{1}) using a fourth-order
Runge-Kutta algorithm in time, and fourth-order centered
difference in space scheme. Our initial condition involves two
oppositely moving dark solitons that collide at the center of the
impurity, with the form,
\begin{equation}
\psi(x,0) = \psi_b(x) \{\cos\varphi_1 \tanh[\cos\varphi_1 (x+x_1)] +
i\sin\varphi_1\} \{\cos\varphi_2 \tanh[\cos\varphi_2 (x-x_2)] -
i\sin\varphi_2\}~,\label{3}
\end{equation}
where $0\leq\varphi_{1,2}<\pi/2$, $x_{1,2}>0$, and
$x_1/x_2=\sin\varphi_1/\sin\varphi_2$. For sufficiently large values
of $x_1$ and $x_2$, Eq.~(\ref{3}) approximately represents a pair of
two dark solitons located at $-x_1$ and $x_2$, with oppositely
moving velocities $\sin\varphi_1$ and $-\sin\varphi_2$. The above
condition for $x_1/x_2$ then ensures that the solitons arrive at the
the center of the impurity concurrently. In our setup, we control a
small difference between $\varphi_1$ and $\varphi_2$, ensuring that
$|(\varphi_2-\varphi_1)/\varphi_1| \leqslant0.15$. Two normalized
integral quantities on each side of the barrier are computed in the
numerical experiments\footnote{It should be noted that a collision
of dark soliton(s) with the barrier always results in pairwise
emission of much smaller dark and anti-dark entities on each side.
For the anti-dark entity, the integration $\int (\psi_b^2-|\psi|^2)
dx<0$, and, the emission of such a negative portion may affect the
mass redistribution within the left and right portions of the
domain. We verified that this effect had no significant bearing on
the reported results.}
\begin{equation}\label{4}
E^-(t) = \frac{\int_{-\infty}^{0}(\psi_b^2-|\psi|^2)dx}
{\int_{-\infty}^{+\infty}(\psi_b^2-|\psi|^2)dx}~~~~~
E^+(t) = \frac{\int_{0}^{+\infty}(\psi_b^2-|\psi|^2)dx}
{\int_{-\infty}^{+\infty}(\psi_b^2-|\psi|^2)dx}~.
\end{equation}
It is easily understood by symmetry that for $\varphi_1 =\varphi_2$
and other parameters chosen the same for both incoming dark
solitons, we obviously obtain $E^-=E^+=0.5$ after collision. We now
consider the case with small asymmetry, and compute a phase
diagram of $E^\pm$ after collision for both slow and fast solitons.
In the simulation, we control $|E_0^\pm-0.5|\leqslant0.03$ [$E^\pm(t=0)$ is denoted by $E_0^\pm$] in order to satisfy the small
difference between the normalized masses of the two incoming dark
solitons; the results are presented in Fig.~\ref{A}.
\\[\intextsep]
\begin{minipage}{\textwidth}
\renewcommand{\captionlabeldelim}{:}
\renewcommand{\figurename}{Figure }
\renewcommand{\captionfont}{ }
\renewcommand{\captionlabelfont}{ }
\vspace{0cm}\centering
\includegraphics[scale=0.6]{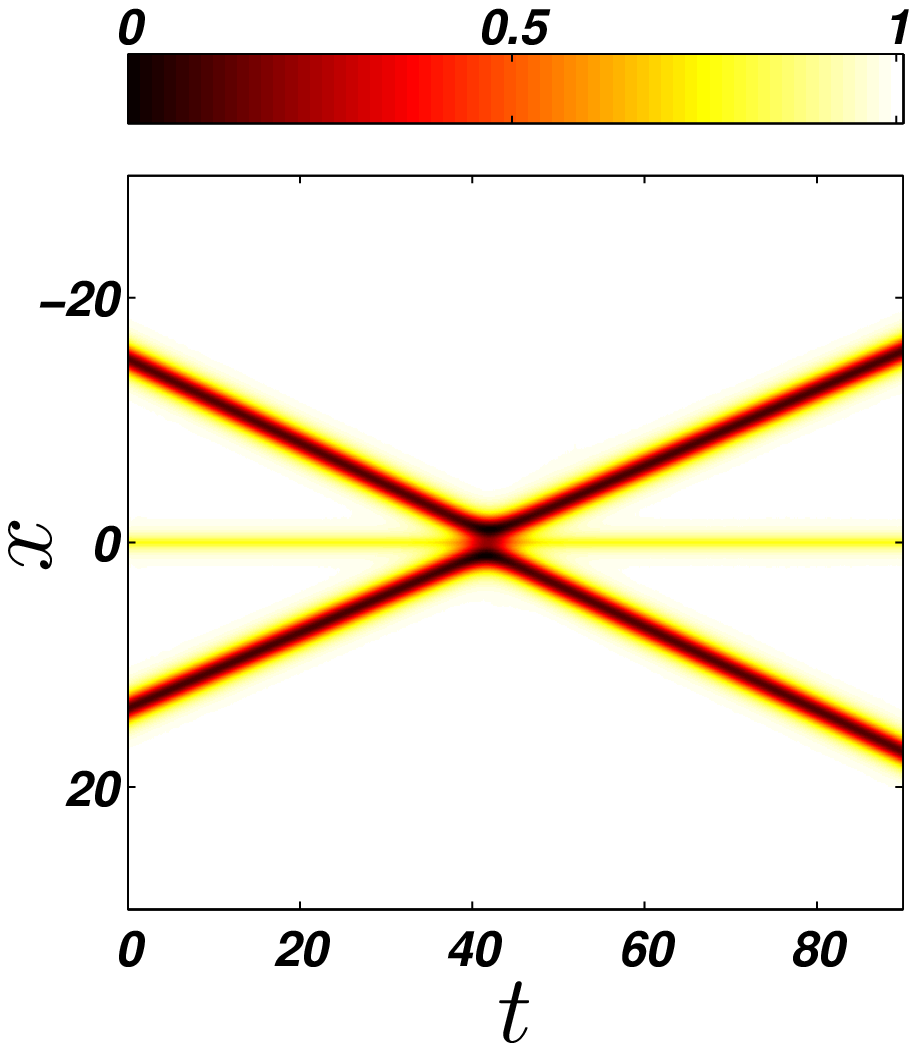}
\hspace{10mm}
\includegraphics[scale=0.6]{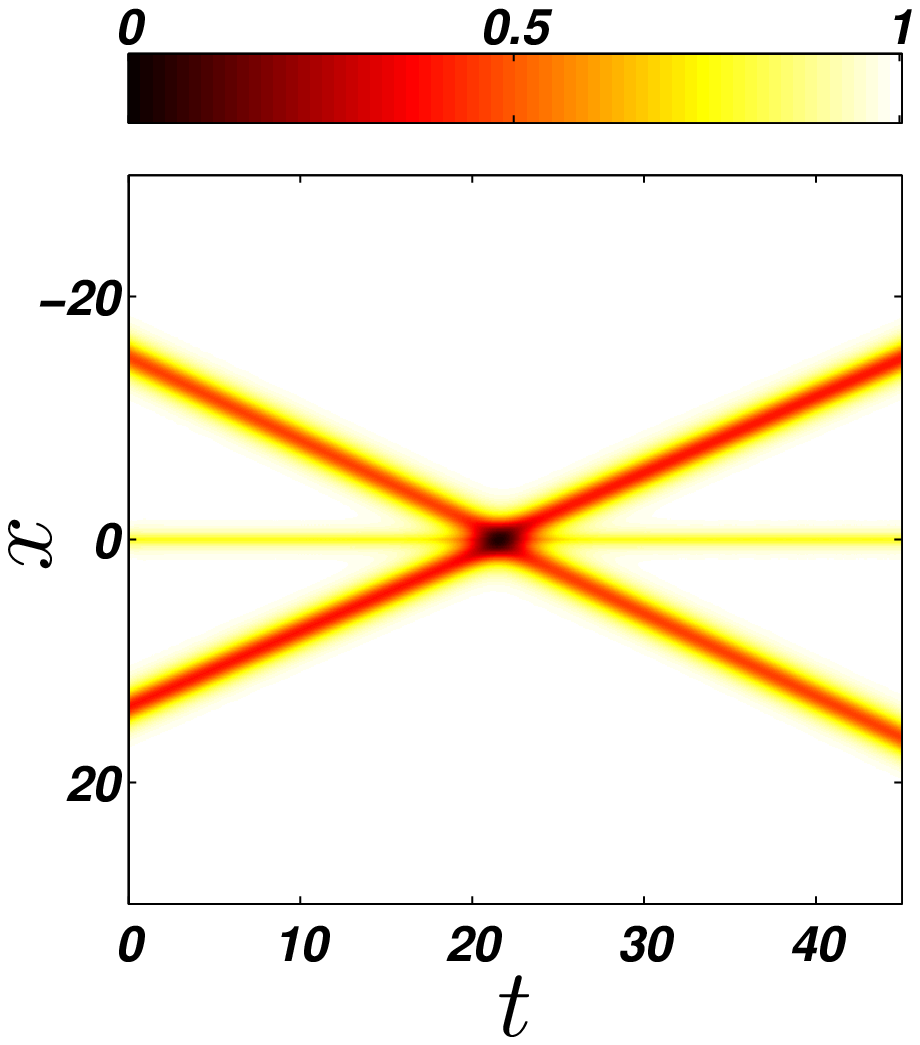}
\figcaption{ Numerical simulation of a two-dark-soliton collision at
center of the impurity. The relevant parameters are $q=0.3$,
$\sigma=0.1$, $\mu=1$, $x_1=15$, and $\varphi_2/\varphi_1=0.90$. (a)
$\varphi_1=0.35$; (b) $\varphi_1=0.75$. }\label{AB}
\end{minipage}
\\[\intextsep]

From this phase diagram, we see that slight initial differences
cannot generate amplified asymmetry in the output. In fact, the
basic behavior hereby is simultaneous reflection (transmission) of
the two incoming solitons by (through) the Gaussian barrier. Two
typical examples are shown in Fig.~\ref{AB}, where the left panel is
for simultaneous reflection, while the right panel is for
simultaneous transmission. 
We vary the parameters for the repulsive impurity ($q$ and
$\sigma$), background ($\psi_0$), and initial soliton position
($x_1$) in the numerical simulation and the results retain features
similar to the one shown above. On the other hand, for an attractive
impurity ($q<0$), the dark soliton (pair) can generally pass through
the barrier, being unable to cause the HOM-analog asymmetry. A
particular case occurs for a pair of slow solitons with slight
asymmetry: after the collision, one soliton is trapped by the
barrier for a short while before it is released to either side of
the barrier. However, such an atypical situation is not included in
our analogue.

\section{Scattering of DB-soliton pairs by impurity}
Given the limited ability of dark solitons to feature
HOM type extrema in transmission/reflection, which can possibly
be partially attributed to their topological character (associated
with a phase slip), we now turn to composite structures featuring
a non-topological (bright) component, namely to dark-bright solitary
waves.
An associated physical system can for example be composed of two different
hyperfine states of the same alkali isotope~\cite{revip}. If this condensate is
confined in a highly anisotropic trap, with the longitudinal frequency
much smaller than the transverse frequency, the mean-field dynamics of
the BEC can be described by the following dimensionless system of
two coupled GP equations,
\begin{subequations}\label{5}
\begin{align}
&\hspace{0mm} i \frac{\partial\psi_1(x,t)}{\partial t} =
-\frac{1}{2} \frac{\partial^2\psi_1(x,t)}{\partial x^2} +
(|\psi_1(x,t)|^2 + |\psi_2(x,t)|^2)\psi_1(x,t)+V_1(x)\psi_1(x,t)~,\label{5a}\\
&\hspace{0mm} i \frac{\partial\psi_2(x,t)}{\partial t} =
-\frac{1}{2} \frac{\partial^2\psi_2(x,t)}{\partial x^2} +
(|\psi_1(x,t)|^2 + |\psi_2(x,t)|^2)\psi_2(x,t) +
V_2(x)\psi_2(x,t)~.\label{5b}
\end{align}
\end{subequations}
Scaling of Eqs.~(\ref{5}) and the relevant physical units can be found
in \cite{Achilleos,Alvarez}. The (ratio of) nonlinearity coefficients
is taken to be unity, which leads the system of Eqs.~(\ref{5}) to
a variant of the well-known
Manakov model~\cite{prinari}.
Such an assumption is consistent with experiments
based on two different hyperfine states of $^{87}$Rb \cite{Mertes,
Hamner,Hoefer,Middelkamp}, where the scattering lengths characterizing
the intra- and inter-component atomic collisions are almost equal\footnote{It should be borne in mind that slight deviations from the limit
especially towards the immiscible side may be responsible for
fundamentally different dynamical evolutions involving phase
separation between the components~\cite{Mertes}. Yet,
DB solitons have been identified as existing on both sides
of this transition~\cite{fotini}.}.
On
the other hand, $V_{1,2}(x)$ represent the normalized external potentials;
in our setup, we consider a localized Gaussian potential (impurity) added in the
component 2 that supports a bright soliton, namely $V_2(x)=
\frac{q}{\sqrt{2\pi}\sigma} e^{-\frac{x^2}{2\sigma^2}}$ and $V_1(x)=0$.
The potential can be generated by off-resonant Gaussian laser beams,
and for a blue- or red-detuned laser beam, the impurity potential
can either repel ($q>0$) or attract ($q<0$) the
atoms of the relevant component of the condensate.

For our analog of the HOM phenomenology,
the Gaussian impurity plays the role of the splitter,
and the DB-soliton pairs with slight asymmetry play the role of photons.
With the boundary conditions $|\psi_1|^2\rightarrow\mu$ and
$|\psi_2|^2\rightarrow0$ as $|x|\rightarrow\infty$, the incoming soliton
pairs (the initial conditions in the simulation) are chosen as the following
form that describes two DB solitons colliding at the center of the impurity:
\begin{subequations}\label{6}
\begin{align}
&\hspace{0mm} \psi_1(x,0) = \sqrt{\mu}
\{\cos\alpha_1\tanh[k_1(x+x_1)] + i\sin\alpha_1\}
\{\cos\alpha_2\tanh[k_2(x-x_2)] - i\sin\alpha_2\}~,\label{6a}\\
&\hspace{0mm} \psi_2(x,0) = A_1 \textrm{sech}[k_1(x+x_1)] e^{i v_1
x} + A_2 \textrm{sech}[k_2(x-x_2)] e^{-i (v_2 x+\Delta)}~,\label{6b}
\end{align}
\end{subequations}
where $\alpha_j$ is the dark soliton's phase angle, $\sqrt{\mu}\cos\alpha_j$
and $A_j$ are the amplitudes of the dark and bright solitons, $k_j$ and
$(-1)^j x_j$ are associated with the inverse width and the initial position
of the DB solitons, and $(-1)^{j-1}v_j$ and $\Delta$ represent the soliton
velocity and a relative phase. These parameters of the DB-soliton pairs
satisfy the following relations:
\begin{subequations}\label{7}
\begin{align}
&\hspace{0mm} k_j^2 + A_j^2 = \mu\cos^2\alpha_j~,\label{7a}\\
&\hspace{0mm} v_j = k_j \tan\alpha_j~~~(j=1,2),\label{7b}\\
&\hspace{0mm} x_1/v_1 = x_2/v_2~.\label{7c}
\end{align}
\end{subequations}
In our analogue, we choose two independent parameters $k_j$ and $v_j$,
and consider nontrivial deviations
between the parameters of the two DB solitons.
In this case, there are two sets of masses, respectively, for
the dark and bright components in order to quantify relevant transfer.
\begin{subequations}\label{8}
\begin{align}
&\hspace{0mm} E_B^- (t) =
\frac{\int_{-\infty}^{0}|\psi_2|^2dx}{\int_{-\infty}^{+\infty}|\psi_2|^2dx}~~~~~E_B^+ (t)
=
\frac{\int_{0}^{+\infty}|\psi_2|^2dx}{\int_{-\infty}^{+\infty}|\psi_2|^2dx}~,\label{8a}\\
&\hspace{0mm} E_D^- (t) =
\frac{\int_{-\infty}^{0}(\mu-|\psi_1|^2)dx}{\int_{-\infty}^{+\infty}(\mu-|\psi_1|^2)dx}~~~~~E_D^+ (t)
=
\frac{\int_{0}^{+\infty}(\mu-|\psi_1|^2)dx}{\int_{-\infty}^{+\infty}(\mu-|\psi_1|^2)dx}. \label{8b}
\end{align}
\end{subequations}
We prescribe these normalized {masses} to feature
small deviations from symmetry (with
$|E_{B,D}^\pm(0)-0.5|\leqslant0.03$ in general). The simulation results
will be systematically presented below (in the following figures, the notations
$\psi_1\rightarrow\psi_D$ and $\psi_2\rightarrow\psi_B$ are commonly used).
\\[\intextsep]
\begin{minipage}{\textwidth}
\renewcommand{\captionlabeldelim}{:}
\renewcommand{\figurename}{Figure }
\renewcommand{\captionfont}{ }
\renewcommand{\captionlabelfont}{ }
\vspace{0cm}\centering
\includegraphics[scale=0.42]{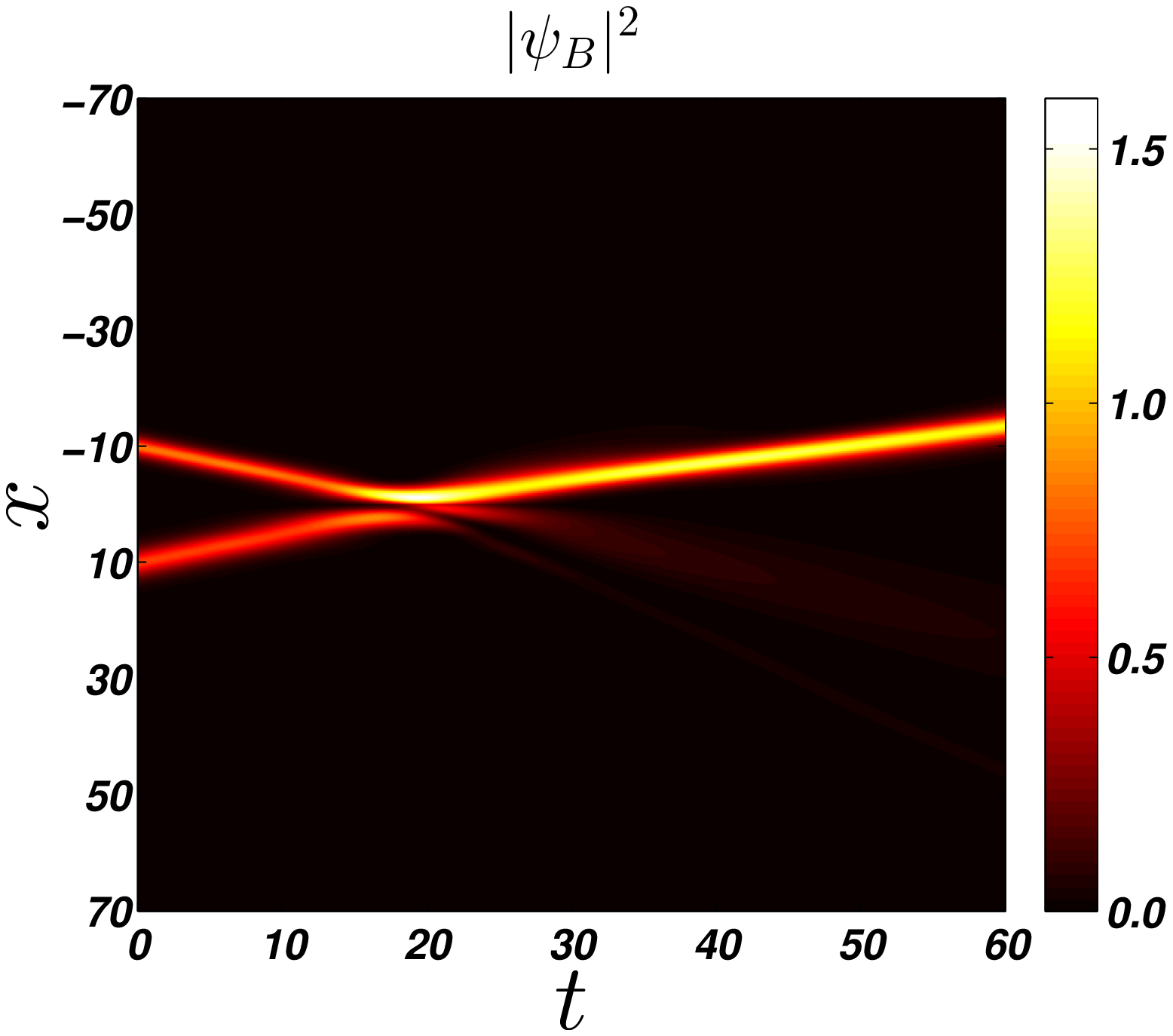}
\includegraphics[scale=0.42]{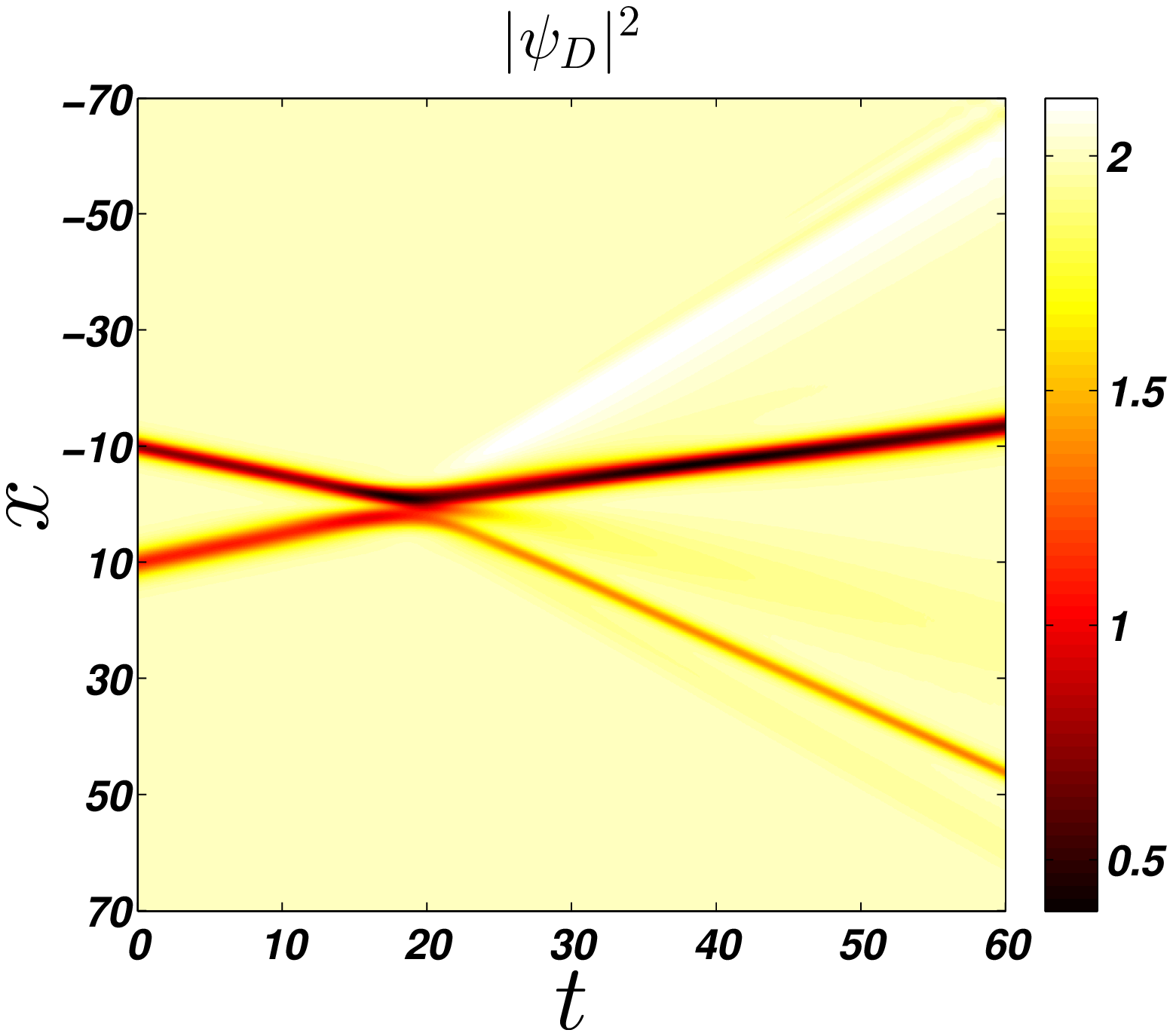}
\figcaption{Numerical simulation of a two-DB-soliton collision at the center
of impurity. The relevant parameters are $q=1$, $\sigma=0.1$, $\mu=2$, $k_1=0.66$,
$k_2=0.46$, $v_1=v_2=0.5$, $x_1=10$, and $\Delta=0$. }\label{B}
\end{minipage}
\\[\intextsep]

Firstly, we consider the case of unequal inverse widths, and
illustrate typical realizations in Fig.~\ref{B}. With the parameters
in this figure, the difference in the inverse widths directly leads
to a very slight asymmetry of $E_D^\pm(0)$ (that is
$|E_D^\pm(0)-0.5|\lesssim1.5\%$), which, after collision at the
impurity, induces a much larger deviation on the normalized masses
($E_D^-\approx60\%$). Figure \ref{C} examines the role of the
difference between $k_1$ and $k_2$ by fixing $k_1$, and varying
$k_2$ in the range of $0.6-1.0$ (ensuring
$|E_D^\pm(0)-0.5|\lesssim3.0\%$), with the results $E_{B,D}^-$
(after collision) shown in this figure. We see that a peak value
occurs for both $E_{B,D}^-$ when $k_2/k_1$ varies in the range,
which means more of the soliton mass (or the normalized mass) is
found on one side. We observe that such a maximum asymmetry for the
bright (non-topological)
component is considerably stronger than that for the dark component. On
the other hand, the situation is almost symmetric as $k_2/k_1$
varies from 1 to higher values (not shown here).
\\[\intextsep]
\begin{minipage}{\textwidth}
\renewcommand{\captionlabeldelim}{:}
\renewcommand{\figurename}{Figure }
\renewcommand{\captionfont}{ }
\renewcommand{\captionlabelfont}{ }
\vspace{0cm}\centering
\includegraphics[scale=0.41]{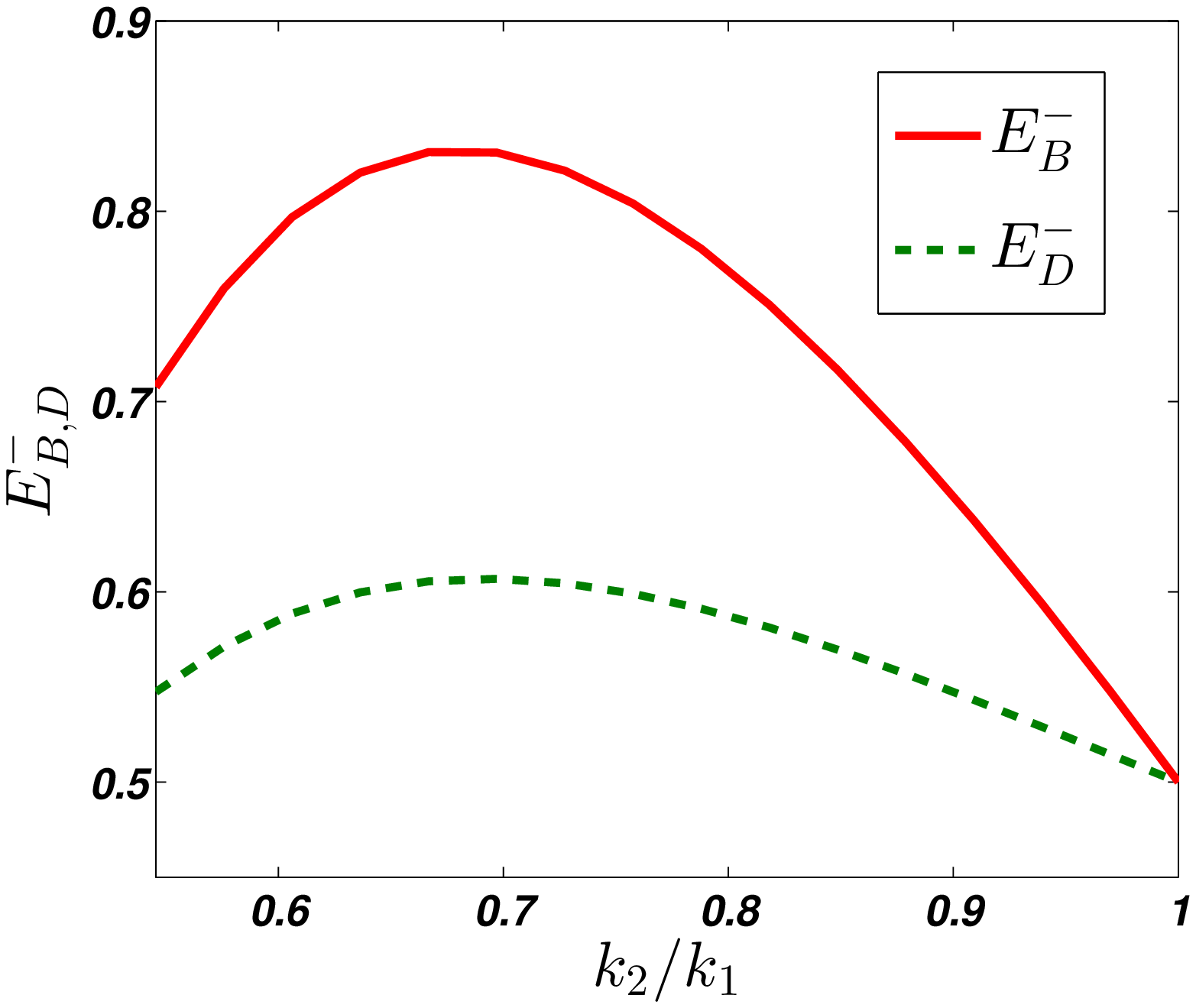}
\figcaption{Plots of $E_{B,D}^-$ as a function of $k_2/k_1$ varying
from 0.6 to 1.0 ($k_1=0.66$). The relevant parameters are $q=1$,
$\sigma=0.1$, $\mu=2$, $v_1=v_2=0.5$, $x_1=10$, and $\Delta=0$.
}\label{C}
\end{minipage}
\\[\intextsep]
\\[\intextsep]
\begin{minipage}{\textwidth}
\renewcommand{\captionlabeldelim}{:}
\renewcommand{\figurename}{Figure }
\renewcommand{\captionfont}{ }
\renewcommand{\captionlabelfont}{ }
\vspace{0cm}\centering
\includegraphics[scale=0.40]{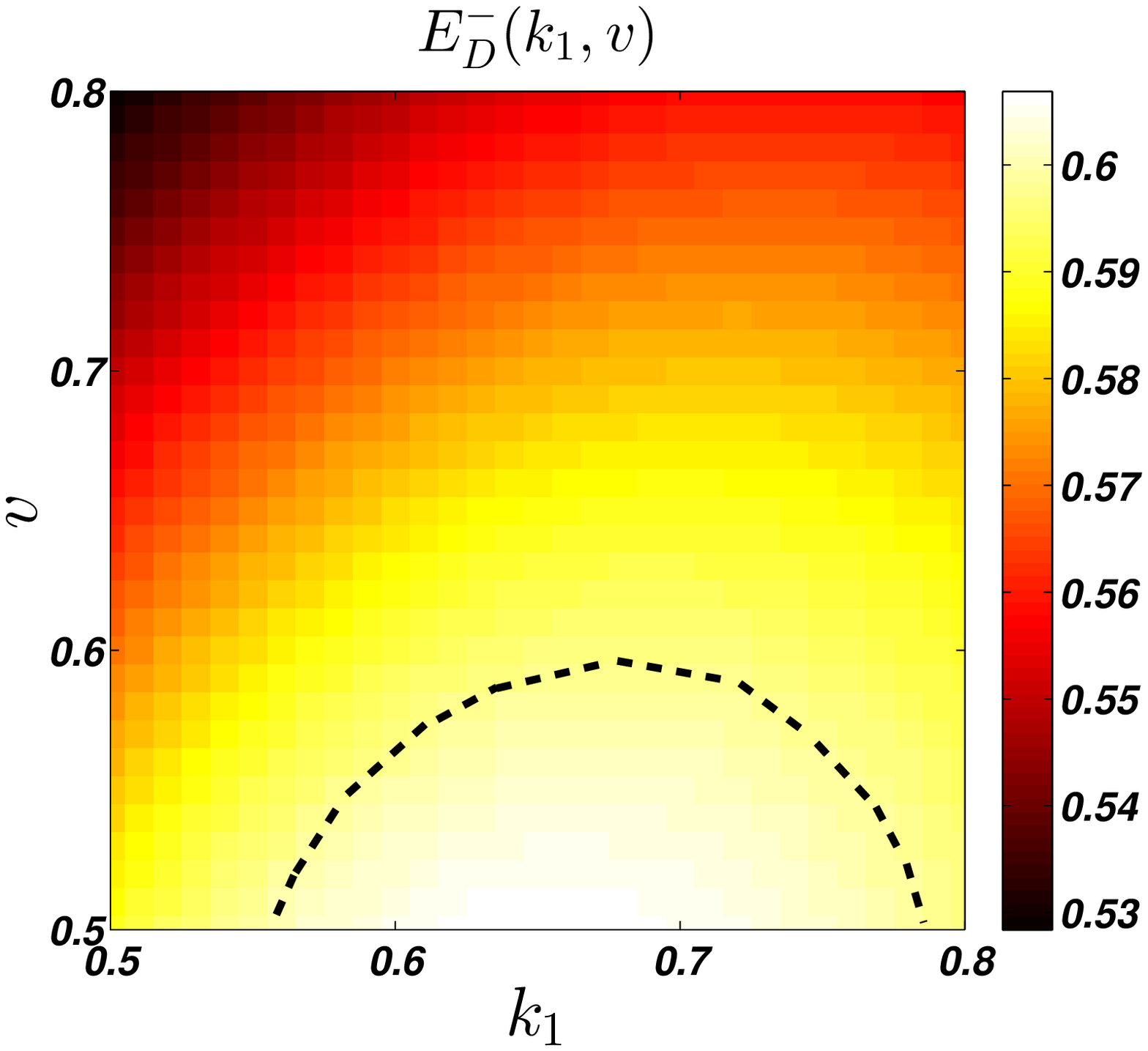}
\includegraphics[scale=0.40]{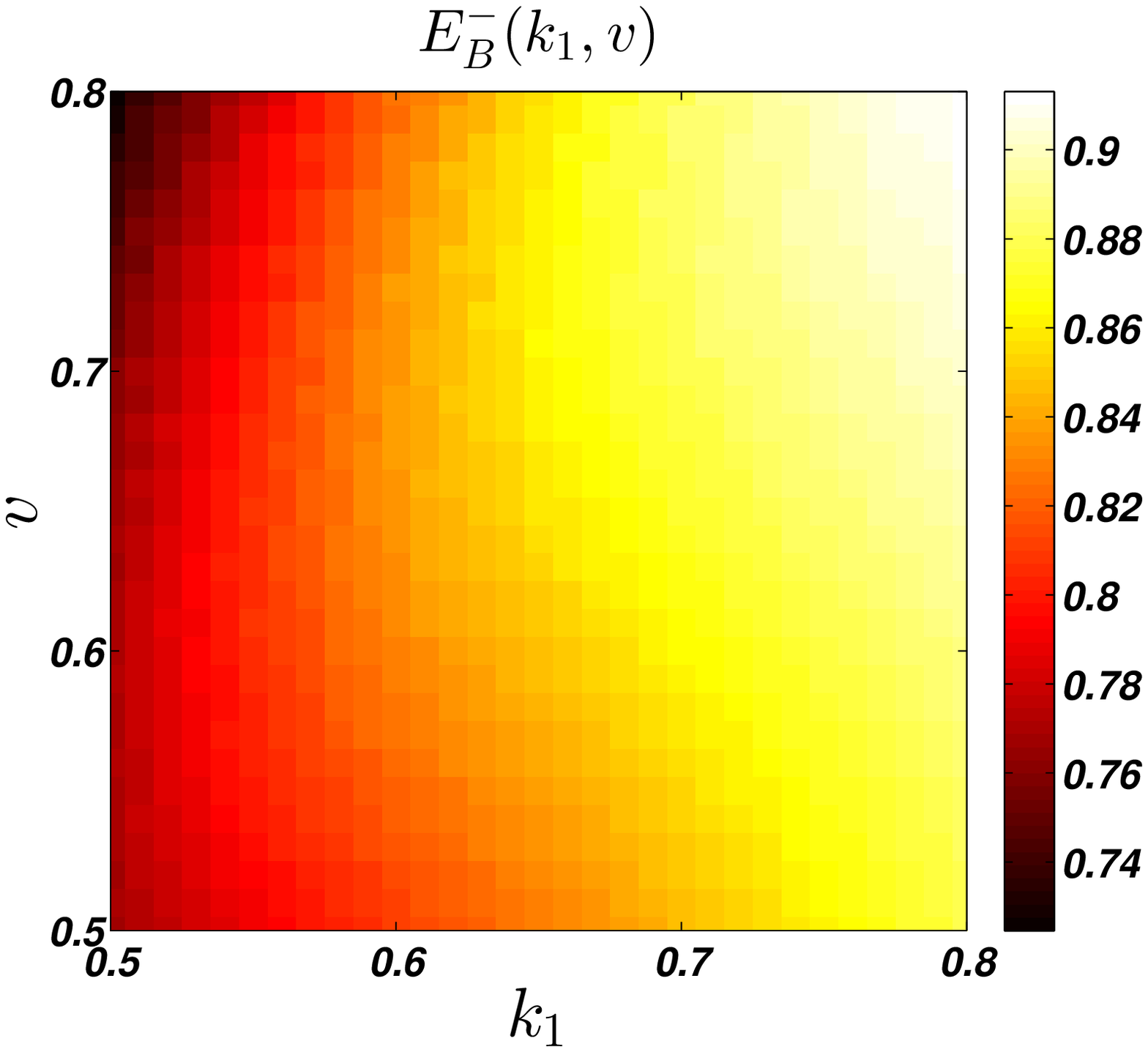}
\figcaption{Two-parameter diagram of the propagation asymmetry
for the dark (left) and bright (right) components of the DB
solitons. The relevant parameters are $q=1$, $\sigma=0.1$, $\mu=2$,
$x_1=10$, and $\Delta=0$. }\label{D}
\end{minipage}
\\[\intextsep]

To further check the dependence of the maximum asymmetry with $k_1$
and $v$ ($v_1=v_2=v$), a phase diagram is presented in Fig.~\ref{D},
where fixing each group of $(k_1,v)$, we vary the value of
$k_2/k_1$, and capture the maximum asymmetry of both $E_{B,D}^-$. It
can be seen that features reminiscent of the HOM asymmetry are
clearly evident for the slow solitons with $k_1$ varying in the range $0.55-0.75$
(the corresponding suitable regime is within the dashed line in the figure).
The maximum asymmetry with $E_D^-\approx60\%$ is induced by much
smaller initial deviation $|E_D^\pm(0)-0.5|\lesssim1.5\%$.
However, the phenomenology is fundamentally more pronounced
in the bright component where it is clear that a
behavior reminiscent of the HOM effect can be classically observed for these non-topological
waves with $E_B^{-}$ exceeding values of $0.9$. A case example of
this is shown in Fig.~\ref{B}.

Another interesting possibility is to
explore the behavior of the DB solitons
for an attractive impurity ($q<0$). We firstly discuss the case
where an asymmetry is induced between $k_1$ and $k_2$ (for typical
parameters, we control $k_2/k_1$ varying in the range of $0.8-1.0$,
keeping $|E_D^\pm(0)-0.5|\lesssim3.0\%$). For $q<0$, a very slight
portion of soliton mass (normalized mass) is trapped by the impurity
after the collision. This hardly influences the
phenomenology, and the integration boundary in (\ref{8}) can be
carefully selected\footnote{For instance, the integration can be
revised as $\int_{-\infty}^0\rightarrow\int_{-\infty}^{-\sigma}$ and
$\int^{+\infty}_0\rightarrow\int^{+\infty}_{\sigma}$ for this
situation.}. We study the dependence of $E_{B,D}^-$ as $k_2/k_1$
varies from $0.8$ to $1.0$, with two groups of results provided in
Fig.~\ref{E} (two fixed values of $k_1$ are chosen). We see that
generally for the dark
(component) solitons
the asymmetric output is more pronounced  with small velocity.
Again, these asymmetries
are much stronger in the non-topological component carrying
the bright structure, rather than in the topological dark solitons.
\\[\intextsep]
\begin{minipage}{\textwidth}
\renewcommand{\captionlabeldelim}{:}
\renewcommand{\figurename}{Figure }
\renewcommand{\captionfont}{ }
\renewcommand{\captionlabelfont}{ }
\vspace{0cm}\centering
\includegraphics[scale=0.37]{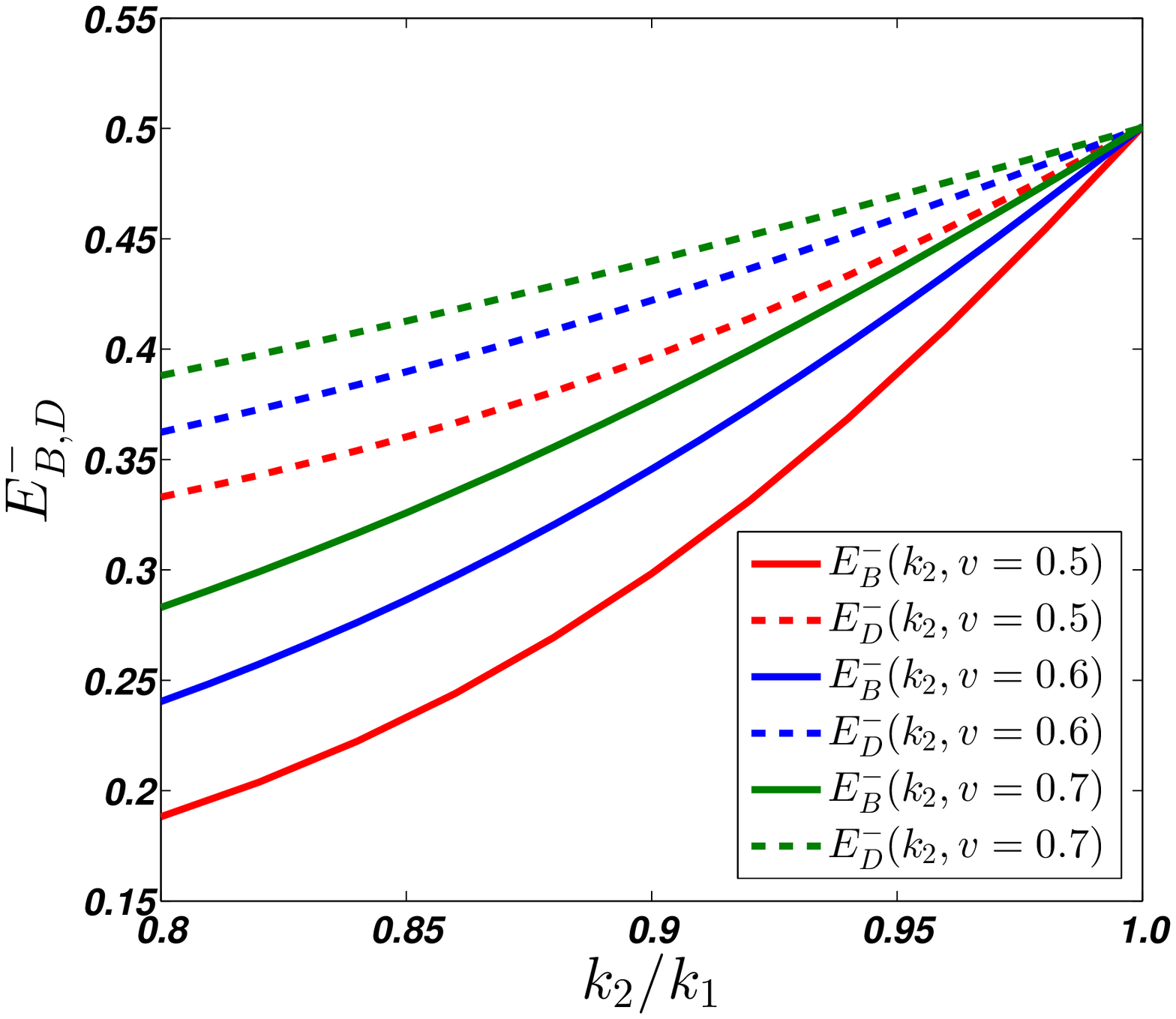}
\includegraphics[scale=0.37]{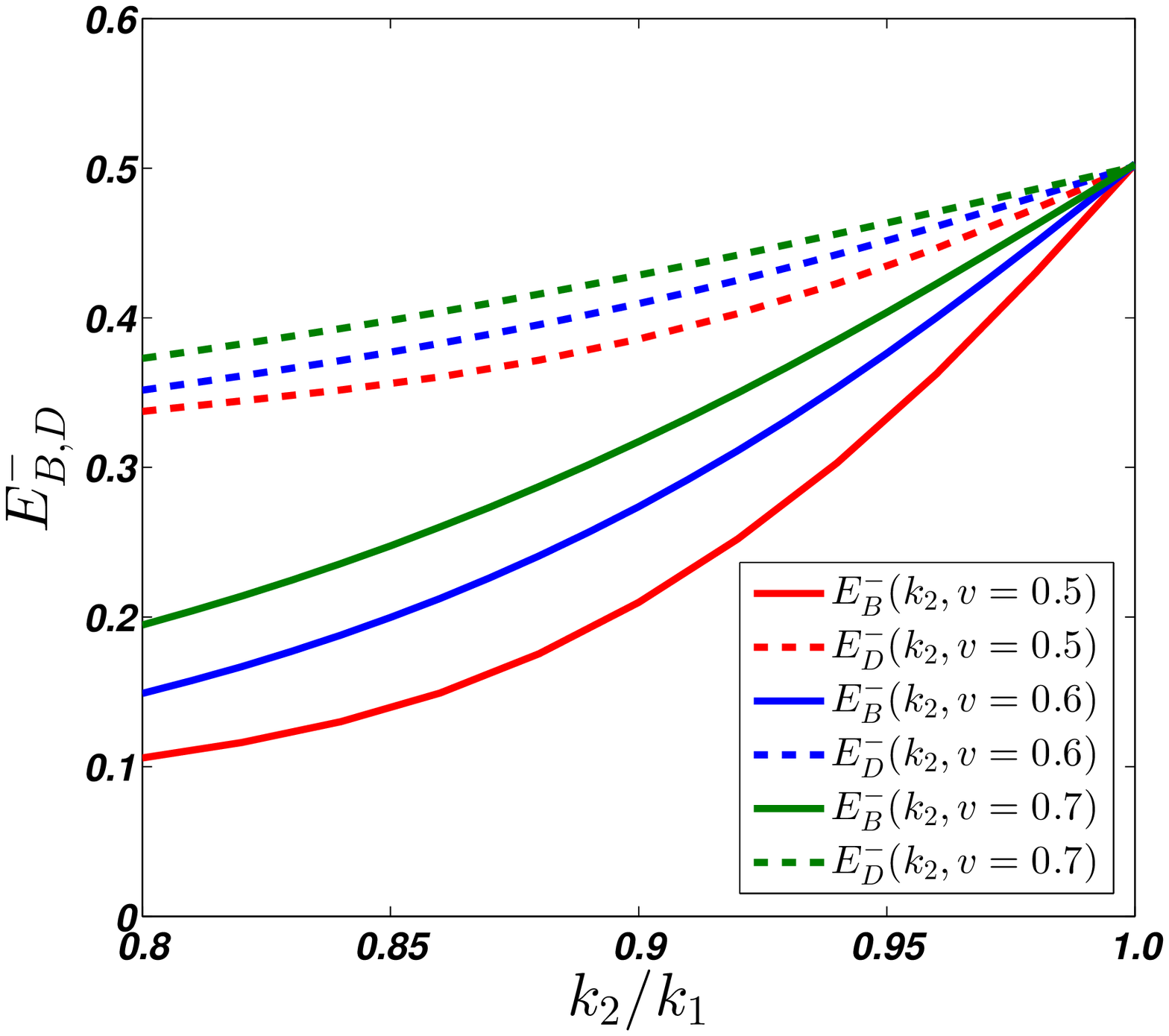}
\figcaption{Plots of $E_{B,D}^-$ as a function of $k_2/k_1$ varying
from 0.8 to 1.0 ($k_1=0.40$ for the left panel, and $k_1=0.50$ for
the right panel). The relevant parameters are $q=-1$, $\sigma=0.1$,
$\mu=2$, $x_1=10$, and $\Delta=0$. }\label{E}
\end{minipage}
\\[\intextsep]

A variation of the subject is that deviations in soliton velocities
may also induce an asymmetry of the collisional output. Since the
deviation $|E_D^\pm(0)-0.5|$ markedly increases with increasing difference between $v_1$ and $v_2$ (setting $k_1=k_2=k$), we
control $|E_D^\pm(0)-0.5|\lesssim3.0\%$ in our simulations. In this situation the function $E_{B,D}^-$ varies with $v_2/v_1$
($v_1$ is fixed) is similar to that of Fig.~\ref{E}. Therefore, we
capture the maximum asymmetry, and draw a two-parameter diagram for the
dependence of the corresponding $E_{B,D}^-$ on $v_1$ and $k$, as
shown in Fig.~\ref{F}. It can be seen that the asymmetric outcome is
more pronounced for the fast-moving solitons for both of the dark
and bright components, with the bright components, as usual,
featuring the most dramatic asymmetry.
This feature is partially different from the
one of Fig.~\ref{D}, where the maximum asymmetry tends to occur for
the narrower solitons, in particular for the slow dark (component)
solitons.
\\[\intextsep]
\begin{minipage}{\textwidth}
\renewcommand{\captionlabeldelim}{:}
\renewcommand{\figurename}{Figure }
\renewcommand{\captionfont}{ }
\renewcommand{\captionlabelfont}{ }
\vspace{0cm}\centering
\includegraphics[scale=0.38]{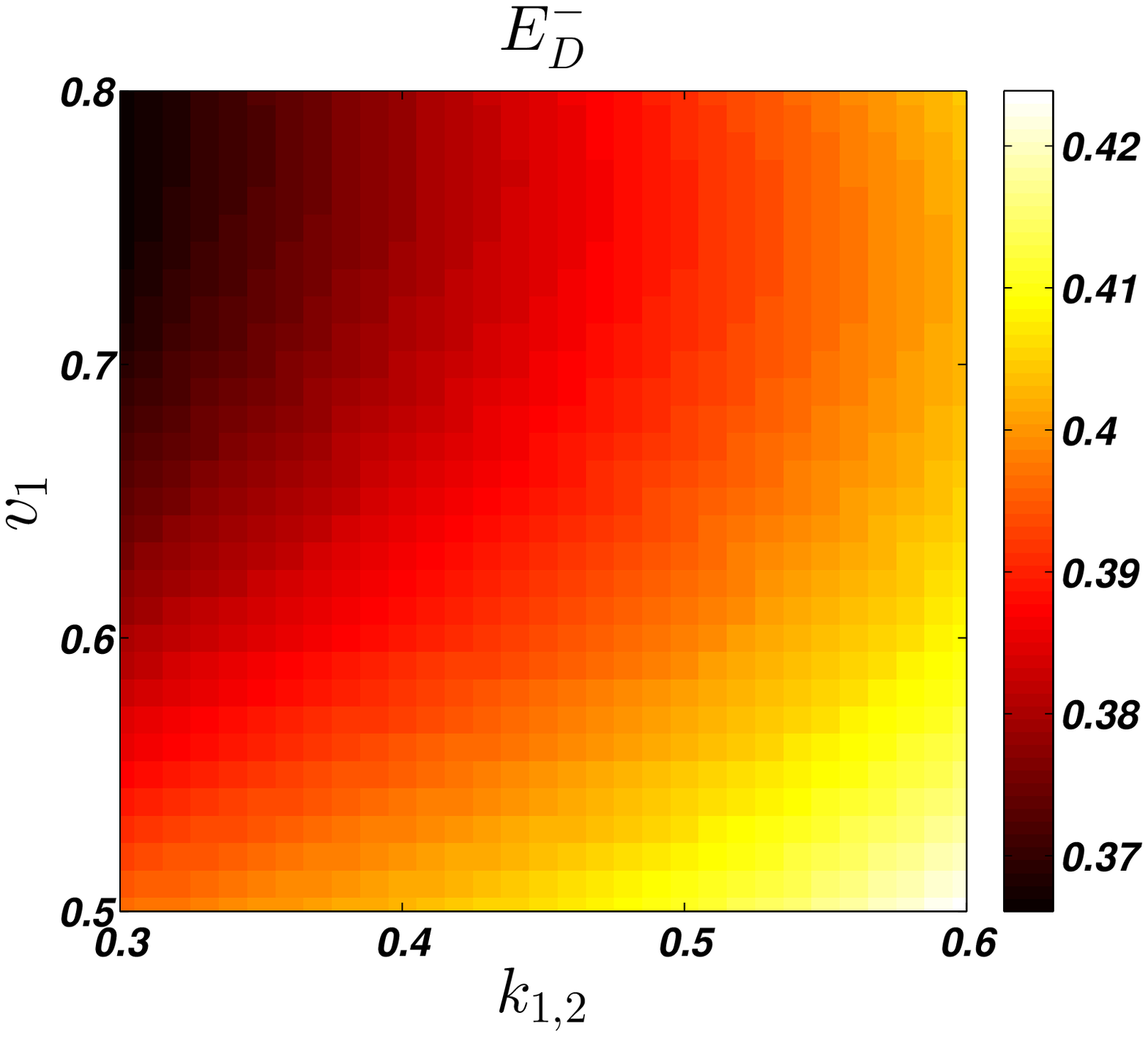}
\includegraphics[scale=0.38]{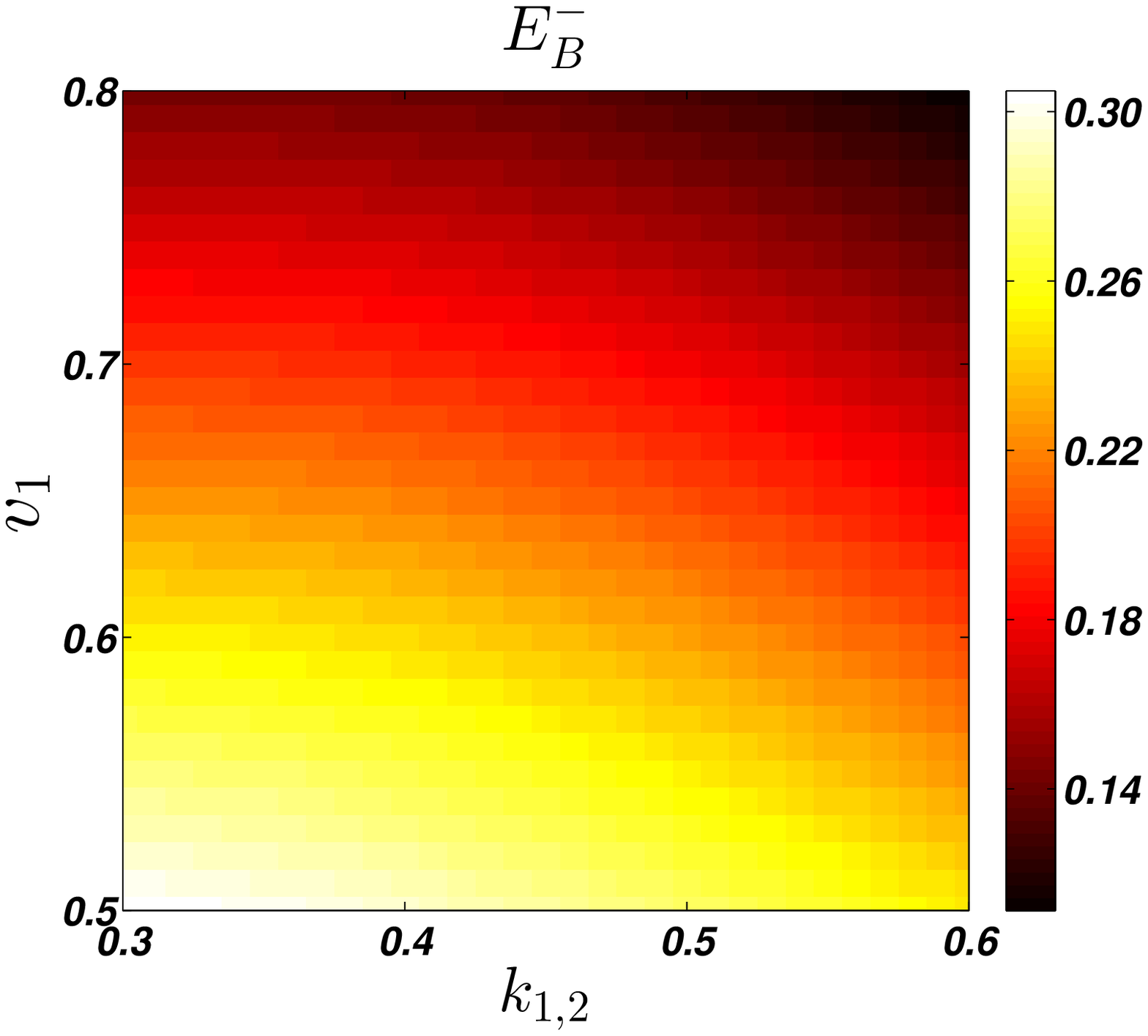}
\figcaption{Two-parameter diagram of the maximum asymmetry for the DB
solitons, again for dark (left) and bright (right) components.
The relevant parameters are $q=1$, $\sigma=0.1$, $\mu=2$,
$x_1=20$, and $\Delta=0$. }\label{F}
\end{minipage}
\\[\intextsep]
\\[\intextsep]
\begin{minipage}{\textwidth}
\renewcommand{\captionlabeldelim}{:}
\renewcommand{\figurename}{Figure}
\renewcommand{\captionfont}{ }
\renewcommand{\captionlabelfont}{ }
\vspace{0cm}\centering
\includegraphics[scale=0.39]{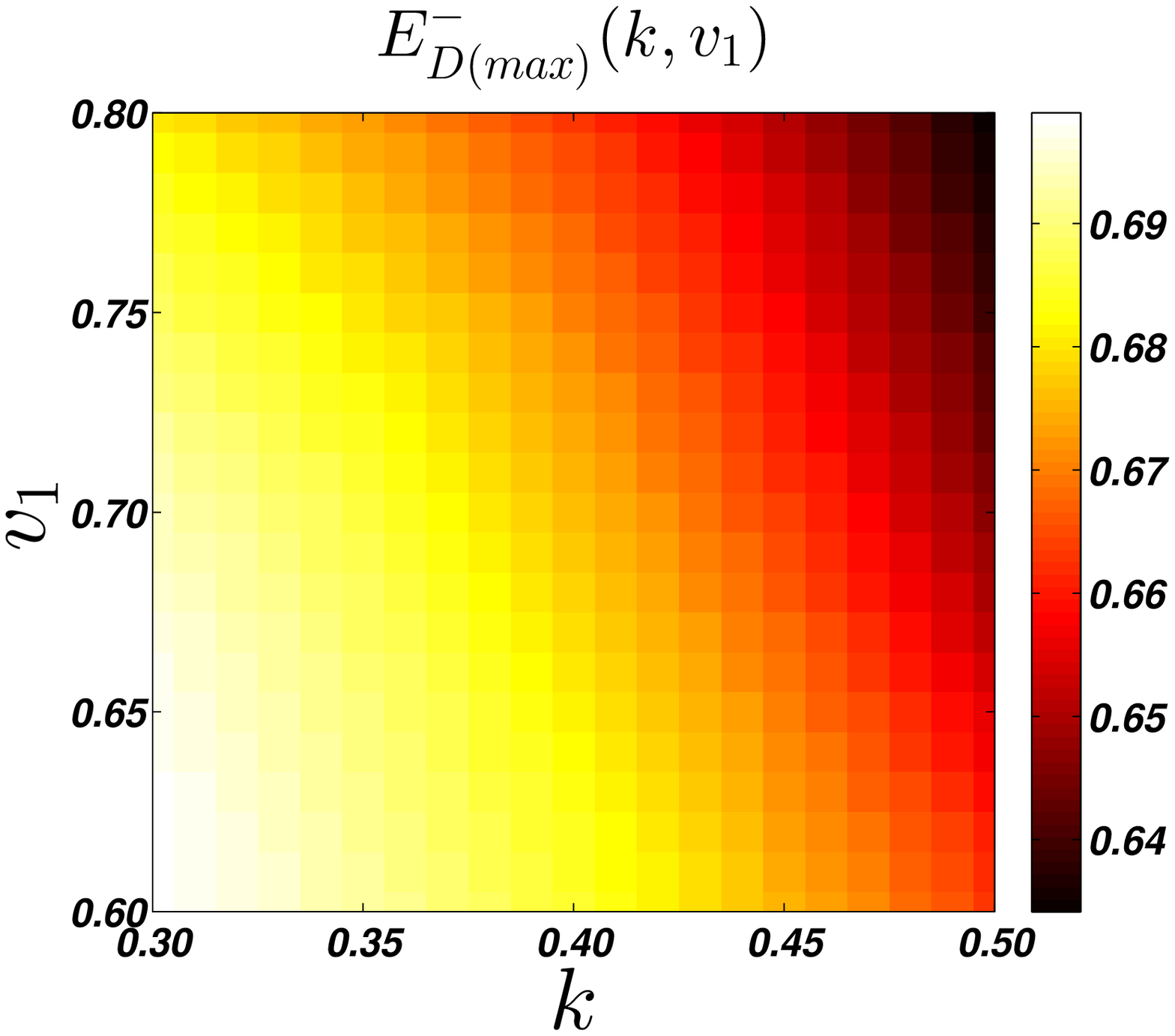}
\figcaption{Two-parameter diagram of the asymmetry for the dark
solitons (component). The relevant parameters are $q=-1$,
$\sigma=0.1$, $\mu=2$, $x_1=20$, and $\Delta=0$. }\label{G}
\end{minipage}
\\[\intextsep]

Also, for the attractive impurity, the deviations of soliton
velocities can induce maximal asymmetry after the soliton collision. Numerical simulations show that this behavior is captured in a
narrow regime of parameters $(k,v_1)$. We study the variation of
$E_{B,D}^-$ as a function of $v_2/v_1$ that varies in the range of
$0.8-1.0$, ensuring $|E_D^\pm(0)-0.5|\lesssim4.0\%$. The functions
are similar to those shown in Fig.~\ref{C}, with a
maximum asymmetry (peak values) as $v_2/v_1$ varies. In the same
way, we draw a phase diagram of the maximum asymmetry for the
parameters $(k,v_1)$, as illustrated in Fig.~\ref{G}. We observe that the outcome after the collision is more asymmetric
for the slower and wider dark (component) solitons.

In addition, we briefly examine the dependence of the asymmetric
output on the starting soliton location $x_1$. We perform
simulations with different selections of $x_1$. The results are
presented in Fig.~\ref{H}. These figures show that the trend of
asymmetry is increasing in general as the location $x_1$ increases.
Figures~\ref{H}(a) and (c) display that the \textit{optimal} point
produces a substantial asymmetry ($k_2/k_1\rightarrow1$ or
$v_2/v_1\rightarrow1$) as $x_1$ increases. Fig.~\ref{H}(b) shows
that for such type of variation, the asymmetry is generally
increasing ($k_2/k_1$ varies in the whole range of $0.8-1.0$) as
$x_1$ is increased.
\\[\intextsep]
\begin{minipage}{\textwidth}
\renewcommand{\captionlabeldelim}{:}
\renewcommand{\figurename}{Figure }
\renewcommand{\captionfont}{ }
\renewcommand{\captionlabelfont}{ }
\vspace{0cm}\centering
\includegraphics[scale=0.3]{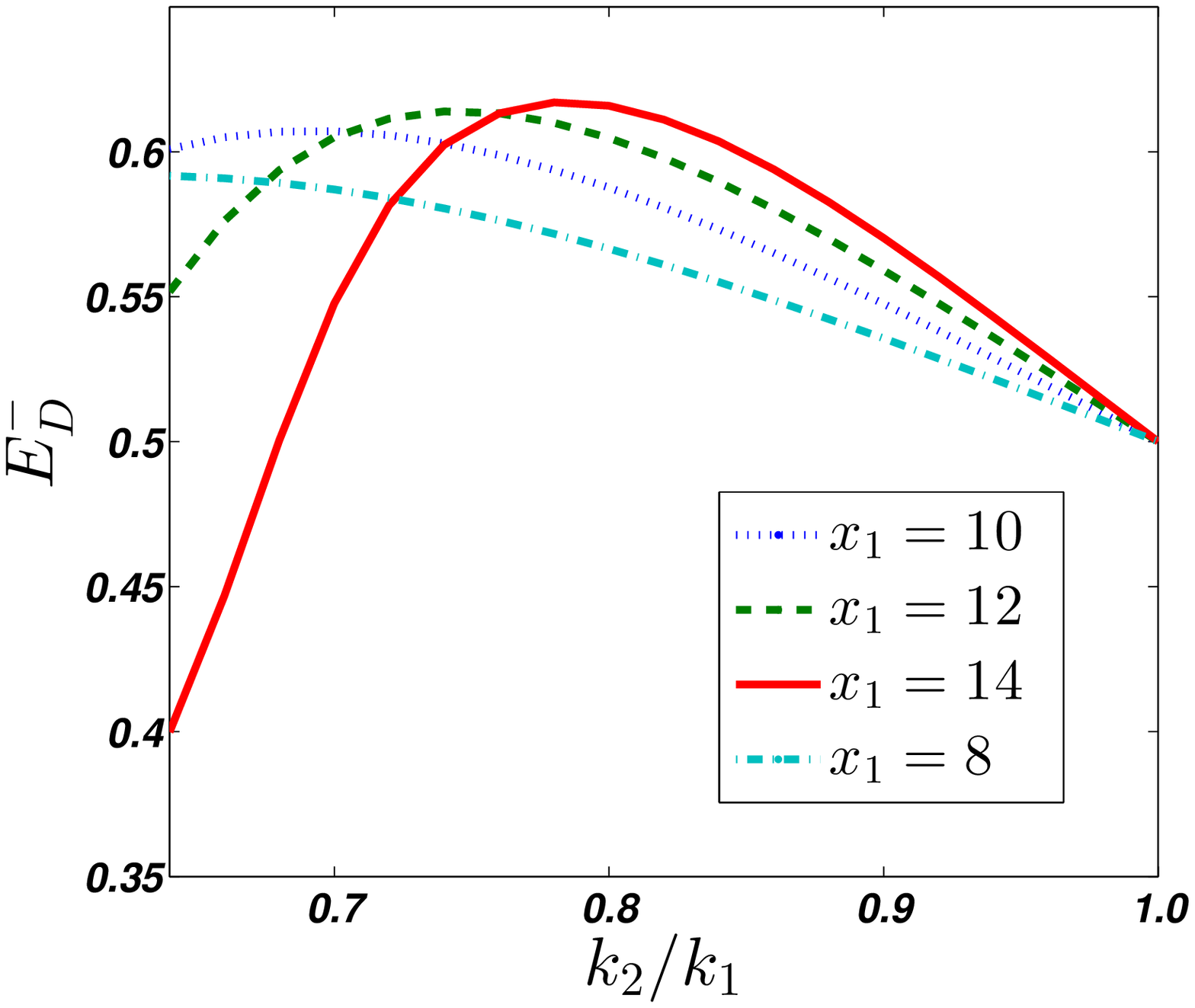}
\includegraphics[scale=0.3]{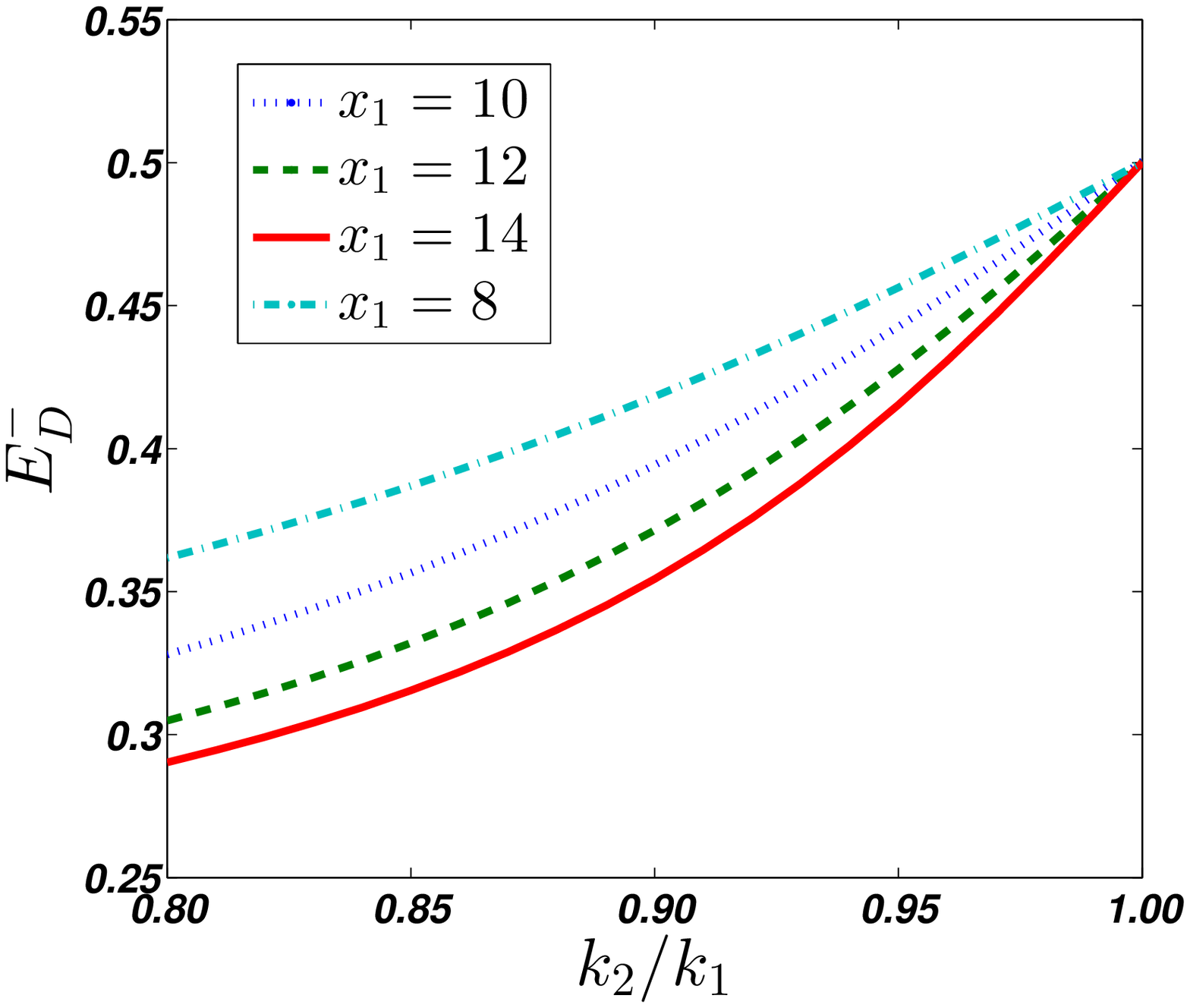}
\includegraphics[scale=0.3]{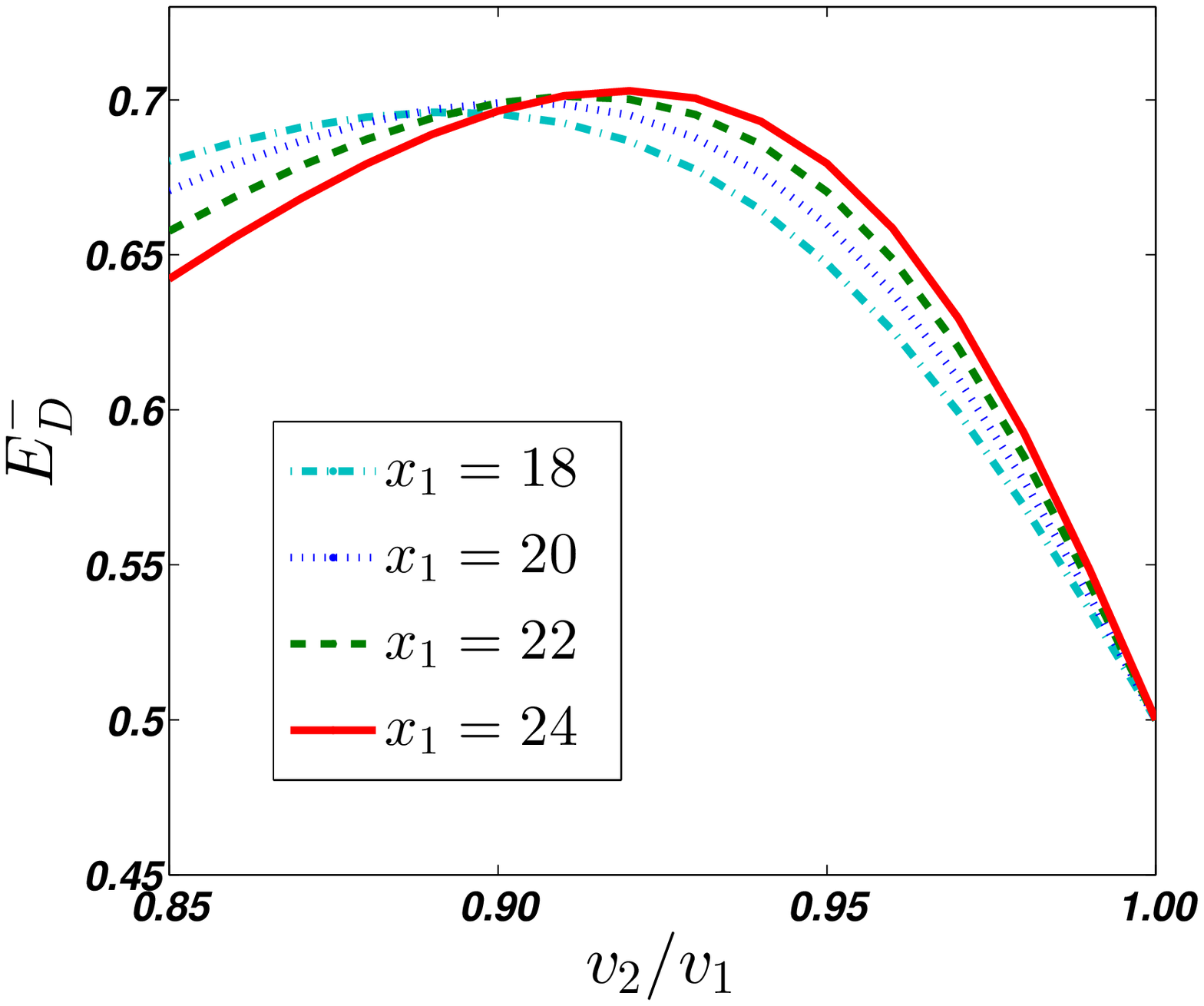}
\figcaption{(a) Plots of $E_{D}^-$ (after collision) as a function
of $k_2/k_1$. The relevant parameters are $q=1$, $\sigma=0.1$,
$\mu=2$, $v_1=v_2=0.5$, $k_1=0.66$, and $\Delta=0$. $x_1$ is equal
to $8$, $10$, $12$, and $14$, respectively. (b) Plots of $E_{D}^-$
as a function of $k_2/k_1$. The relevant parameters are $q=-1$,
$\sigma=0.1$, $\mu=2$, $v_1=v_2=0.5$, $k_1=0.40$, and $\Delta=0$.
$x_1$ is equal to $8$, $10$, $12$, and $14$, respectively. (c) Plots
of $E_{D}^-$ as a function of $v_2/v_1$. The relevant parameters are
$q=-1$, $\sigma=0.1$, $\mu=2$, $k_1=k_2=0.3$, $v_1=0.60$, and
$\Delta=0$. $x_1$ is equal to $18$, $20$, $22$, and $24$,
respectively. }\label{H}
\end{minipage}
\\[\intextsep]

\section{Scattering of ring DB-soliton pairs by impurity}
In this section, we will briefly extend the asymmetric collision to
the 2D case, and illustrate a first example with the ring
DB solitons~\cite{stockhofe}.
We consider the evolution of the two-component BEC very near zero
temperature governed by the following coupled GP equations with
external potential (the two-dimensional Manakov model),
\begin{subequations}\label{9}
\begin{align}
&\hspace{0mm} i \frac{\partial\psi_D}{\partial t} =
-\frac{1}{2}\nabla^2\psi_D +
(|\psi_D|^2 + |\psi_B|^2)\psi_D+V_1(r)\psi_D~,\label{9a}\\
&\hspace{0mm} i \frac{\partial\psi_B}{\partial t} =
-\frac{1}{2}\nabla^2\psi_B + (|\psi_D|^2 + |\psi_B|^2)\psi_B +
V_2(r)\psi_B~,\label{9b}
\end{align}
\end{subequations}
where $\nabla^2 = \frac{\partial^2}{\partial x^2} +
\frac{\partial^2}{\partial y^2}$ and $r^2 = x^2 + y^2$. In order to
study the asymmetric interaction of a ring DB-soliton pair
with a localized ring-shaped impurity, we set the external
potentials as
\begin{equation}
V_1(r)=0~,~~~V_2(r)= \frac{q}{\sqrt{2\pi}\sigma}
e^{-\frac{(r-r_0)^2}{2\sigma^2}}~,\label{10}
\end{equation}
where the ring impurity is localized at $r=r_0$. In simulations, the
initial condition used to integrate Eqs.~(\ref{9}) has the same form
as (\ref{6}), whereby $k_j(x-x_j)$ is replaced by $k_j(r-r_j)$, in
which $r_j$ is the initial ring soliton radius, and relations of
other parameters are similar to (\ref{6}) and (\ref{7}). We
demonstrate a realization in Fig.~\ref{I}, where the ring DB-soliton
pairs collide at time $t\approx10$. It can be seen that the
asymmetric outcome after collision is still valid for the 2D ring DB
solitons. In particular, in this example the inner ring ends
up carrying the majority of the relevant non-topological component
mass, while the outer one is nearly ``extinct'' in its bright
component. This serves to illustrate that there should be
intriguing analogies to the HOM-type phenomenology in higher
dimensions, including possibly ones involving vorticity-bearing
structures, that are worth exploring and comparing/contrasting
with the one-dimensional case.
\\[\intextsep]
\begin{minipage}{\textwidth}
\renewcommand{\captionlabeldelim}{: }
\renewcommand{\figurename}{Figure }
\renewcommand{\captionfont}{ }
\renewcommand{\captionlabelfont}{ }
\vspace{0cm}\centering
\includegraphics[scale=0.45]{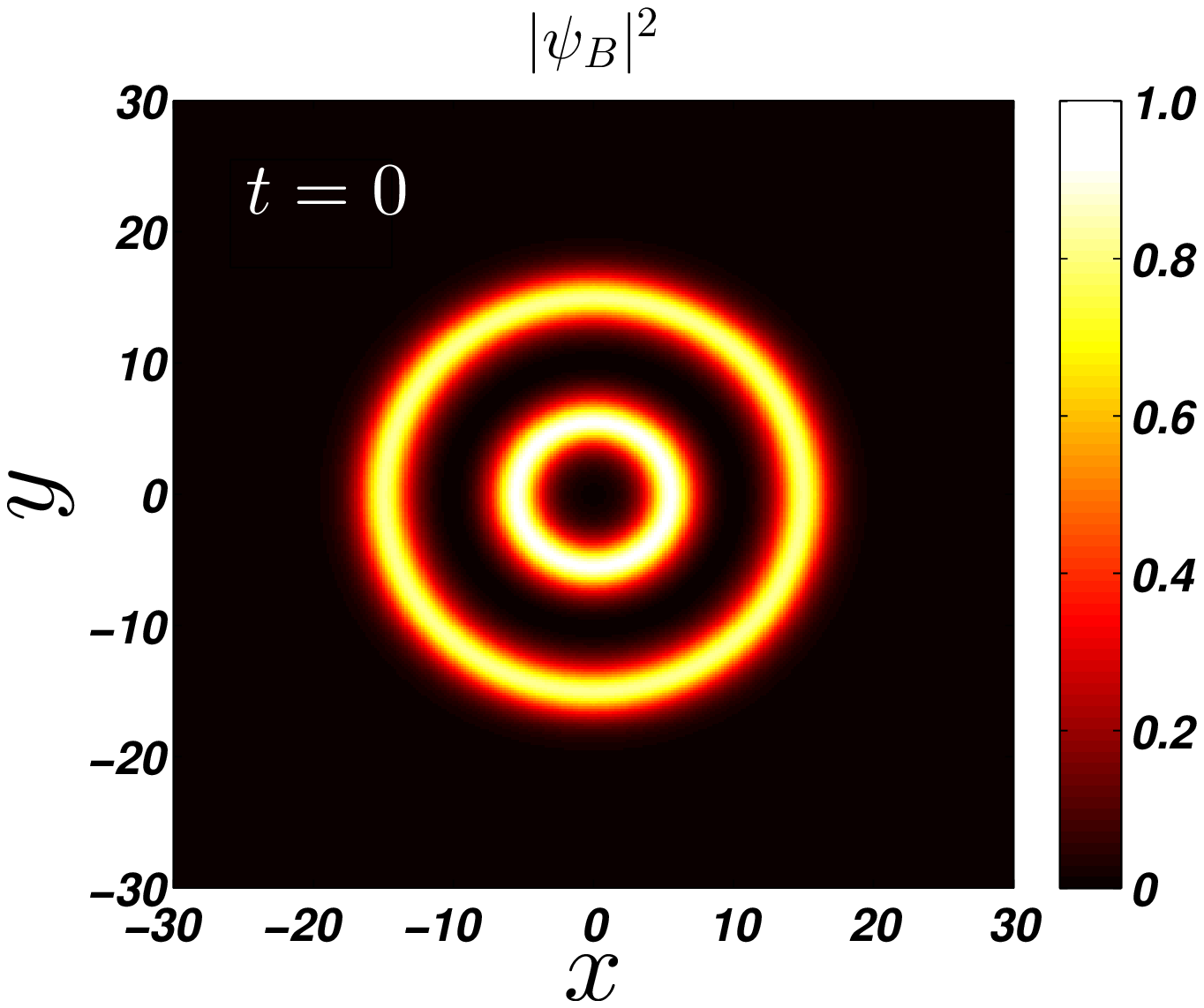}
\includegraphics[scale=0.45]{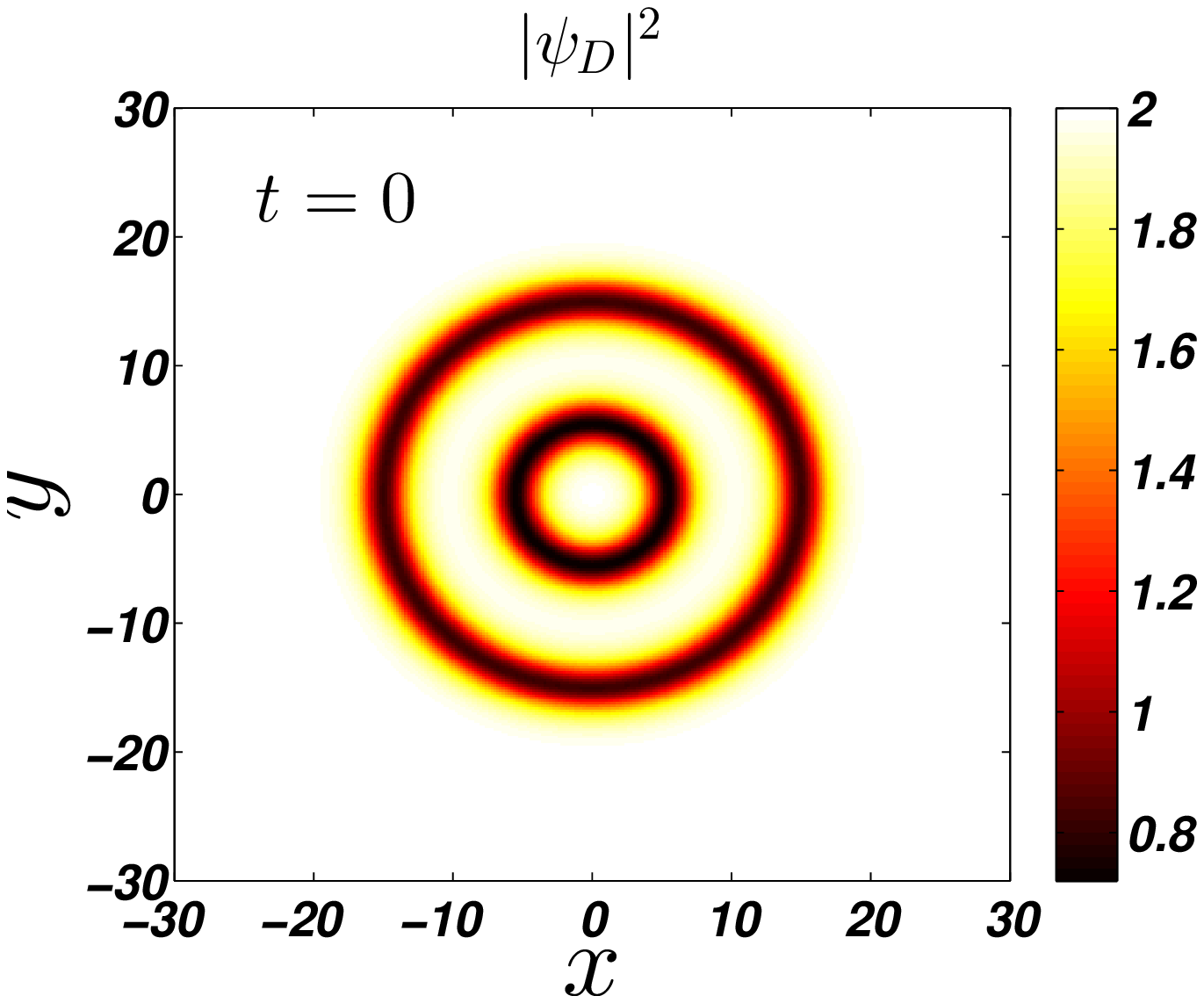}\\
\vspace{5mm}
\includegraphics[scale=0.45]{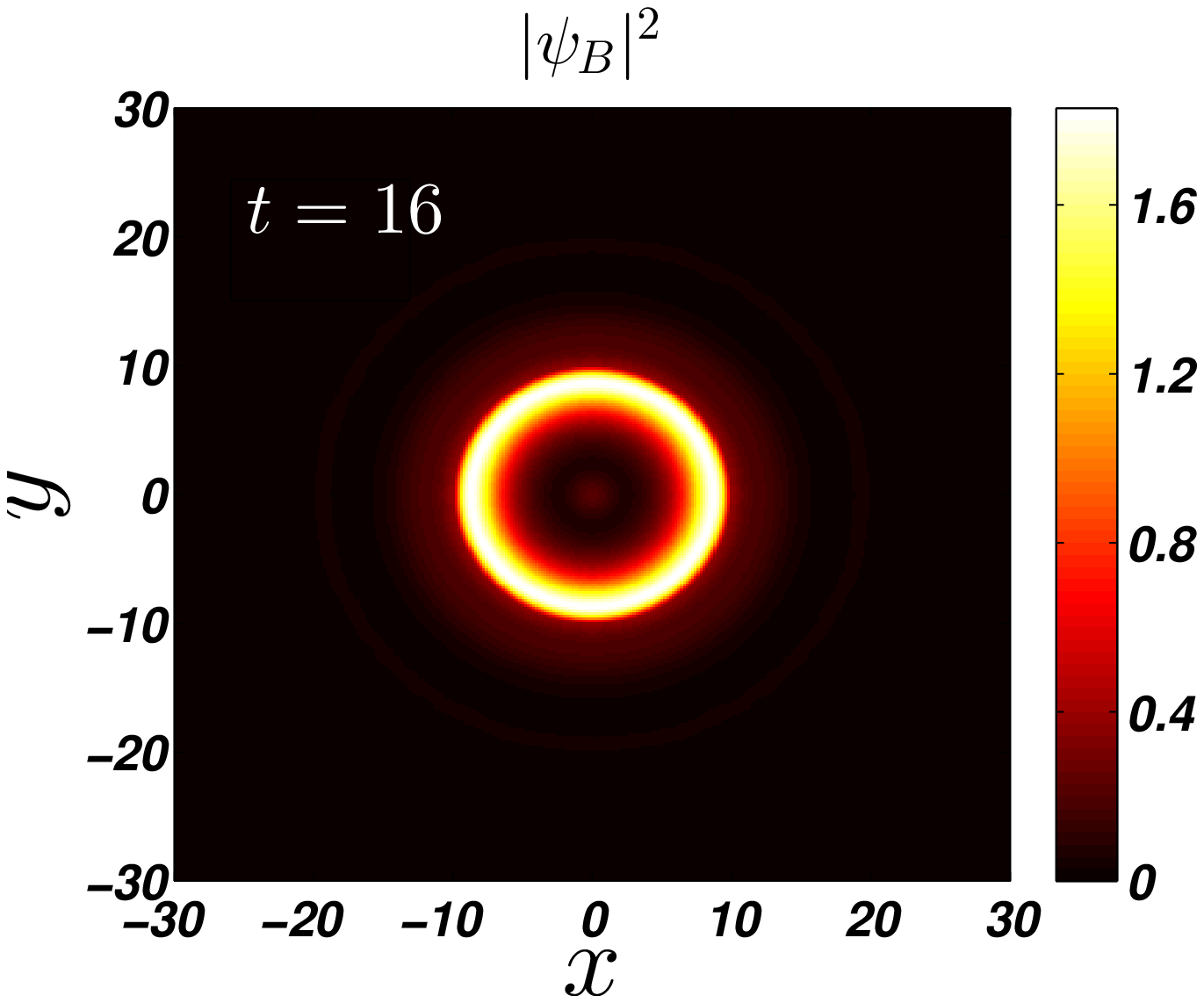}
\includegraphics[scale=0.45]{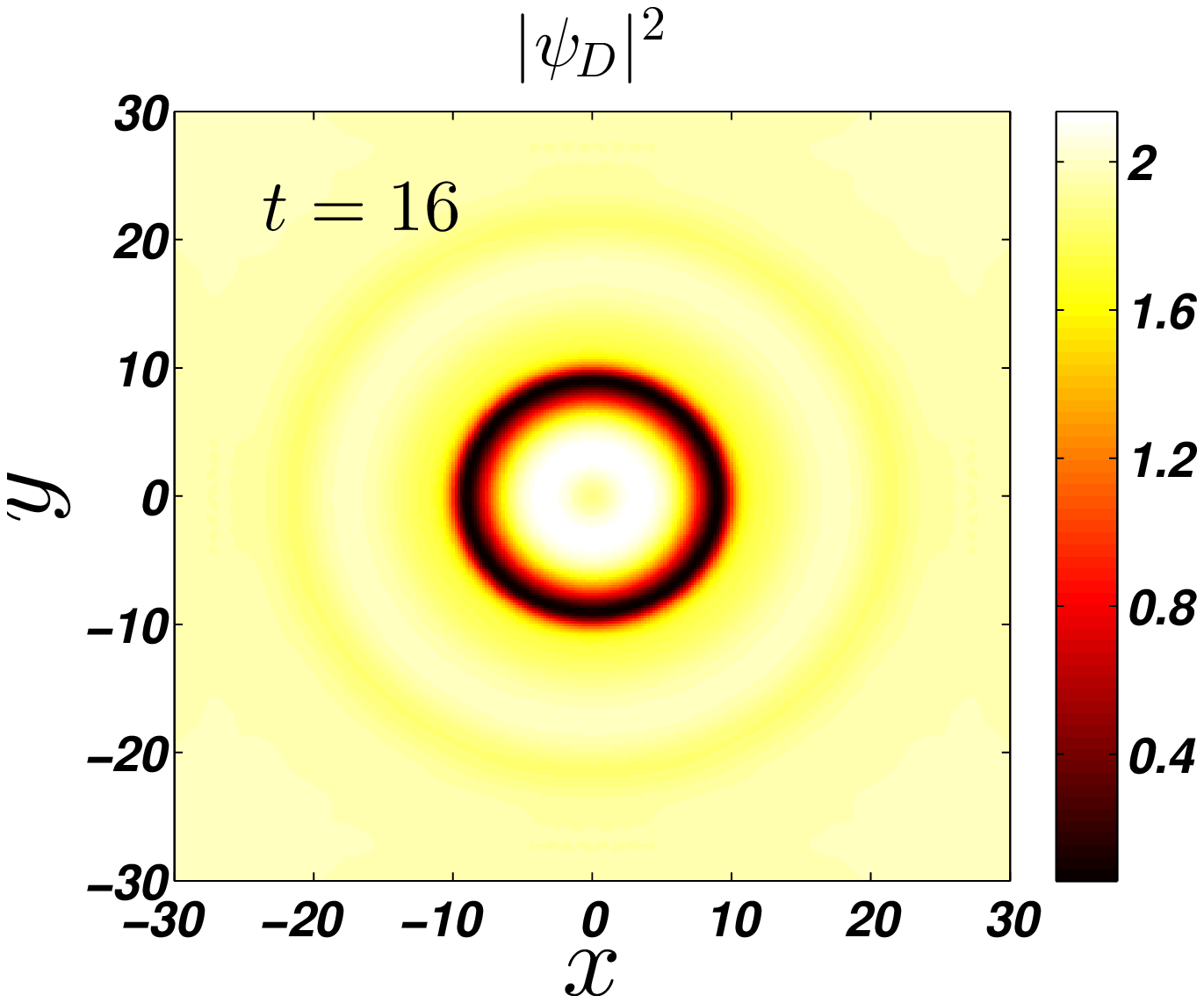}
\figcaption{Numerical simulation of a ring DB soliton collision at
the center of the ring-shaped impurity. The relevant parameters are
$q=1$, $\sigma=0.1$, $\mu=2$, $k_1=k_2=0.60$, $v_1=0.50$,
$v_2=0.45$, $r_1=5$, $r_0=10$, and $\Delta=0$. }\label{I}
\end{minipage}
\\[\intextsep]

\section{Conclusions and Future Challenges}

In the present work we explored the phenomenology of a classical
wave analogue motivated by the Hong-Ou-Mandel effect.
Instead of using
photons and their quantum interference with a beam splitter,
we considered wave-like excitations in a repulsive bosonic
gas described at the mean-field level by a Gross-Pitaevskii
equation. The waves (the interfering entities) were either
dark solitons or dark-bright solitons. The role of the beam
splitter was played by an external Gaussian beam.
 Contrary to our
earlier findings for the potential of bright solitons
to exhibit very sensitive interference patterns
reminiscent of the HOM effect, dark solitons seemed
far less efficient in exhibiting such an effect.
This may arguably be due to their topological character.
This, in turn, led us to explore multi-component dark-bright
entities where the non-topological component is symbiotic
to the topological one, i.e., supported by the dark component
as an effective trapping potential to the bright component.
In this case, the results were far more promising leading the
bright component in one of the waves possibly to nearly
complete extinction, depending on the velocity and width
parameters of the incoming waves. Finally, a proof-of-principle
example was shown for the two-dimensional case of ring dark-bright
solitary waves, where the same phenomenology persisted
in the presence of the curvature associated with the ring-like
excitations.

While this is a first step in this promising direction of research,
numerous additional studies emerge as relevant for future work.
On the one hand, both for the bright and the dark case a quantitative
understanding of the interference phenomenology and how it differs
in the presence of a potential
from the integrable (simply phase shifting) phenomenology of the
integrable cubic nonlinear Schr{\"o}dinger model would be a crucial
contribution to this theme. Generally, the theme of higher dimensional
explorations that we touch upon here is an especially interesting
one. In the bright soliton (focusing or attractive) case, one
can envision for example two solitary waves that are subcritical (or
close to critical) which upon such a collision may become
supercritical in mass and feature collapse rather than
their individual tendency towards dispersion. In the repulsive/defocusing
nonlinearity scenario, understanding the quantitative details of
how curvature affects the picture through ring DB collisions,
or how the presence of vortices (and the interaction of
vortex-bright solitary waves~\cite{revip}) modifies the
mass redistribution are important steps towards a deeper
understanding of the role of dimensionality. In the case of
vortices, it does not escape us that their topological nature
(coupled to the absence of one parameter families of such solutions
for a given background density -- contrary to what is the case
with dark solitons) suggests very minor mass redistributions in
the component bearing the vorticity.
However, mass redistribution is certainly possible and
relevant to explore in the non-topological
component. In the latter, it has been shown to occur even on the basis of
stability properties and tunneling phenomena
alone, rather than collision-induced exchanges~\cite{pola}.
These questions are
currently under consideration and will be reported in future
publications.

\section{Appendix}
To describe a dark soliton on top of the background with impurity,
we write the solution of Eq.~(\ref{1}) in the form,
\begin{equation}
\psi(x,t) = \psi_b(x) e^{-i \psi_0^2 t} \phi(x,t)~,\label{A1}
\end{equation}
with $\phi(x,t)$ chosen as
\begin{equation}
\phi(x,t) = \cos(\varphi) \tanh[\cos(\varphi)(x-x_0)] +
i\sin(\varphi)~,\label{A2}
\end{equation}
where $\varphi$ is a slowly varying function of $t$, and $x_0 =
\int_0^t \sin(\varphi)d\tau$. Following the adiabatic perturbation
approach \cite{Frantzeskakis1,Kivshar}, $\phi(x,t)$ satisfies the
following perturbed equation (assume $\psi_0=1$ without loss of generality),
\begin{equation}
i\frac{\partial\phi}{\partial t} + \frac{1}{2}
\frac{\partial^2\phi}{\partial x^2} - (|\phi|^2-1)\phi =
P(\phi)~,\label{A3}
\end{equation}
where the perturbation $P(\phi)$ has the form
\begin{equation}
P(\phi) = \frac{2H_-(x)}{-4+H_+(x)}\frac{\partial\phi}{\partial x} -
\frac{1}{2}H_+(x) (|\phi|^2-1)\phi~,\label{A4}
\end{equation}
where
\begin{equation}
H_{\pm}(x) = q e^{2\sigma^2} \left[ e^{-2x}
\textmd{erfc}\left(\frac{-x+2\sigma^2}{\sqrt{2}\sigma} \right) \pm
e^{2x} \textmd{erfc}\left(\frac{x+2\sigma^2}{\sqrt{2}\sigma} \right)
\right]~.\label{A5}
\end{equation}
As shown in \cite{Frantzeskakis1,Kivshar}, the evolution equation for
$\varphi(t)$ can be derived as
\begin{equation}
\frac{\partial\varphi}{\partial t} =
\frac{1}{2\cos^2(\varphi)\sin(\varphi)} \textmd{Re} \left[
\int_{-\infty}^{+\infty} P(\phi) \frac{\partial\phi^*}{\partial t}
dx\right]~.\label{A6}
\end{equation}
Substituting (\ref{A4}) into (\ref{A6}), and assuming $\varphi$ to
be a small quantity [i.e., $\sin(\varphi)\approx\varphi$ and
$\cos(\varphi)\approx1-\varphi^2/2$], we obtain the following result
by neglecting the higher-order terms
\begin{equation}
(1-A) \frac{\partial\varphi}{\partial t} = B~,\label{A7}
\end{equation}
where
\begin{subequations}\label{A8}
\begin{align}
&\hspace{0mm} A = \int_{-\infty}^{+\infty} \frac{1}{4} q
e^{2\sigma^2} H_+(x) \textmd{sech}^4(\xi) [1-\xi\tanh(\xi)] dx \notag\\
&\hspace{10mm} + \int_{-\infty}^{+\infty} \frac{\textmd{sech}^4(\xi)
\left[2- q e^{-2x+2\sigma^2}
\textmd{erfc}\left(\frac{-x+2\sigma^2}{\sqrt{2}\sigma}\right)\right]
[\sinh(2\xi)+2\xi]}{-4+q e^{2\sigma^2} H_+(x)} dx~,\label{A8a}\\
&\hspace{0mm} B = \int_{-\infty}^{+\infty} \textmd{sech}^4(\xi)
\left[1-
\frac{1}{4} q e^{2\sigma^2} H_+(x) \tanh(\xi) \right]dx \notag\\
&\hspace{10mm} + \int_{-\infty}^{+\infty} \frac{\textmd{sech}^4(\xi)
\left[4- 2q e^{-2x+2\sigma^2}
\textmd{erfc}\left(\frac{-x+2\sigma^2}{\sqrt{2}\sigma}\right)\right]}{-4+q
e^{2\sigma^2} H_+(x)} dx~,\label{A8b}
\end{align}
\end{subequations}
where $\xi=x-x_0$. Numerically evaluating the integrals of
(\ref{A8}), and considering the effective particle approach for
$x_0$, we can write the effective potential where the soliton
center moves in
\begin{equation}
U(x) = -\int_{\infty}^{x} \frac{d^{2}x_0}{dt^2} dx_0 \approx
-\int_{\infty}^{x} \frac{\partial\varphi}{\partial t}
dx_0~.\label{A9}
\end{equation}
For the repulsive impurity, we can obtain a critical value
$\varphi_c$ that the effective kinetic energy equals
to the height of the effective potential, i.e.,
\begin{equation}
\frac{1}{2} \sin^2(\varphi_c) = U_{max}(x)~.\label{A10}
\end{equation}
Theoretically speaking, when $\varphi>\varphi_c$, the dark soliton
transmits the impurity barrier; otherwise, when $\varphi<\varphi_c$,
the soliton is reflected by the barrier. We perform direct
simulations of Eq.~(\ref{1}) (the numerical method is the same as in
the main content) to find a sequence of values $\varphi_c$, and make
a comparison with (\ref{A10}), as shown in Fig.~\ref{a1}. The
results accord well when $q$ is small ($\varphi_c$ is small as
well), which is reasonable under the assumption of our effective
potential approach.
\\[\intextsep]
\begin{minipage}{\textwidth}
\renewcommand{\captionlabeldelim}{: }
\renewcommand{\figurename}{Figure }
\renewcommand{\captionfont}{ }
\renewcommand{\captionlabelfont}{ }
\vspace{0cm}\centering
\includegraphics[scale=0.60]{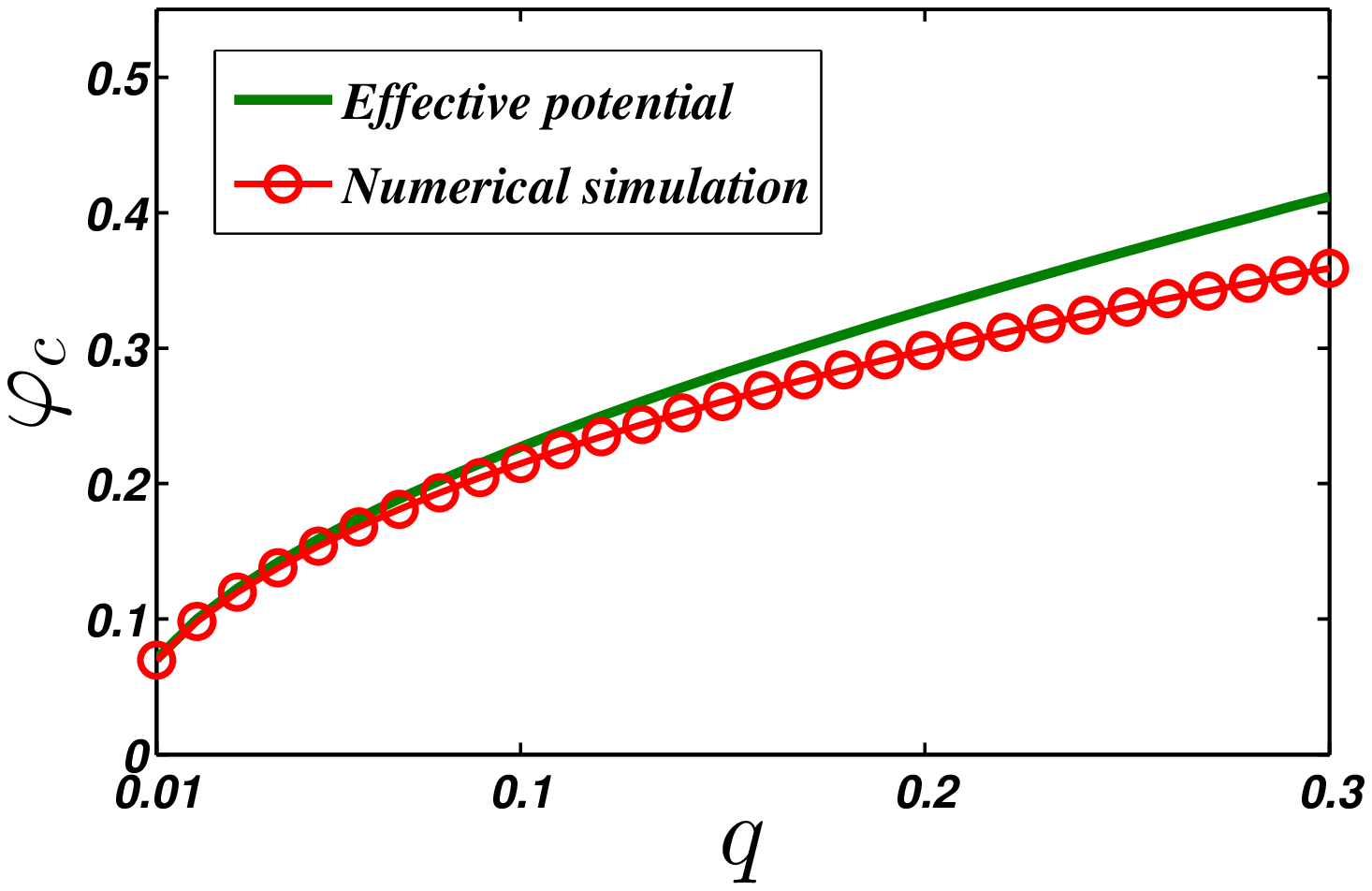}
\figcaption{Comparison of $\varphi_c$ between the effective potential
approach (\ref{A10}) and direct simulation of Eq.~(\ref{1}) ($\sigma=0.1$).
}\label{a1}
\end{minipage}
\\[\intextsep]

\end{document}